\documentclass[a4,usenatbib]{mn2e}
\newcommand{\FRI}{FR{-\small I}}
\newcommand{\FRII}{FR{-\small II}}
\usepackage{graphicx}

\def \etal {\rm ~{\it \etal},~}

\def\apj {{\it Ap.~J.}}

\def\apjs {{\it Ap.~J.\ Suppl.}}
\def\aj {{\it A.~J.}}

\def\aap {{\it Astr.~Ap.}}

\def\japa {{\it J. Ap.\ Astr.}}

\def\mnras {{\it MNRAS}}

\def\pasa {{\it Publications of the Astronomical Society of Australia}}

\title[Radio Galaxy Zoo]{Radio Galaxy Zoo: host galaxies and radio morphologies derived from visual inspection}
\author[J.~K.~Banfield et al.]{J.~K.~Banfield$^{1,2,3}$, O.~I.~Wong$^{4}$, K.~W.~Willett${^5}$, R.~P.~Norris$^{1}$, L.~Rudnick$^{5}$,
\newauthor S.~S.~Shabala$^{6}$, B.~D.~Simmons$^{7}$, C.~Snyder$^{8}$, A.~Garon$^{5}$, N.~Seymour${^9}$, E.~Middelberg$^{10}$,
\newauthor H.~Andernach$^{11}$, C.~J.~Lintott$^{7}$, K.~Jacob$^{5}$, A.~D.~Kapi\'{n}ska$^{4,3}$, M.~Y.~Mao$^{12}$,
\newauthor K.~L.~Masters$^{13,14}$, M.~J.~Jarvis$^{7,15}$, K.~Schawinski$^{16}$, E.~Paget$^{8}$, R.~Simpson$^{7}$,
\newauthor H.-R.~Kl\"ockner$^{17}$, S.~Bamford$^{18}$, T.~Burchell$^{19}$, K.~E.~Chow$^{1}$, G.~Cotter$^{7}$, L. Fortson$^{5}$,
\newauthor I.~Heywood$^{1,20}$,  T.~W.~Jones$^{5}$, S.~Kaviraj$^{21}$, \'A.~R.~L\'opez-S\'anchez$^{22,23}$, W.~P.~Maksym$^{24}$,
\newauthor K.~Polsterer$^{25}$, K.~Borden$^{8}$, R.~P.~Hollow$^{1}$,   L.~Whyte$^{8}$ \newauthor\\
$^{1}$CSIRO Astronomy and Space Science, Australia Telescope National Facility, PO Box 76, Epping, NSW 1710, Australia\\
$^{2}$Research School of Astronomy and Astrophysics, Australian National University, Weston Creek, ACT 2611, Australia\\
$^{3}$ARC Centre of Excellence for All-Sky Astrophysics (CAASTRO)\\
$^{4}$International Centre for Radio Astronomy Research-M468, The University of Western Australia, 35 Stirling Hwy, Crawley, WA 6009, Australia\\
$^{5}$School of Physics and Astronomy, University of Minnesota, 116 Church St. SE, Minneapolis, MN 55455, USA\\
$^{6}$School of Physical Sciences, University of Tasmania, Private Bag 37, Hobart, Tasmania 7001, Australia\\
$^{7}$Oxford Astrophysics, Denys Wilkinson Building, Keble Road, Oxford OX1 3RH, UK\\
$^{8}$Adler Planetarium, 1300 S Lake Shore Dr, Chicago, IL 60605, USA\\
$^{9}$International Centre for Radio Astronomy Research, Curtin University, Perth, Australia\\
$^{10}$Astronomisches Institut, Ruhr-Universit\"at, Universit\"atsstr. 150, 44801 Bochum, Germany\\
$^{11}$Departamento de Astronom\'ia, DCNE, Universidad de Guanajuato, Apdo.\ Postal 144, CP 36000, Guanajuato, Gto., Mexico\\
$^{12}$Joint Institute for Radio Astronomy (JIVE), Postbus 2, 7990 AA Dwingeloo, The Netherlands\\
$^{13}$Institute of Cosmology \& Gravitation, University of Portsmouth, Dennis Sciama Building, PO1 3FX, UK\\
$^{14}$South East Physics Network (SEPNet), {\tt{http://www.sepnet.ac.uk}}\\
$^{15}$Department of Physics, University of the Western Cape, Private Bag X17, Bellville 7535, South Africa\\
$^{16}$Institute for Astronomy, Department of Physics, ETH Z\"urich, Wolfgang-Pauli-Strasse 27, CH-8093 Z\"urich, Switzerland\\
$^{17}$Max-Planck Institut f\"ur Radioastronomie, Auf dem H\"ugel 69, D-53121 Bonn, Germany\\
$^{18}$School of Physics and Astronomy, University of Nottingham, Nottingham NG7 2RD, UK\\
$^{19}$National Radio Astronomy Observatory, PO Box O, Socorro, NM 87801, USA\\
$^{20}$Department of Physics and Electronics, Rhodes University, PO Box 94, Grahamstown 6140, South Africa\\
$^{21}$Centre for Astrophysics Research, University of Hertfordshire, College Lane, Hatfield, Herts, AL10 9AB, UK\\
$^{22}$Australian Astronomical Observatory, PO Box 915, North Ryde, NSW 1670, Australia \\
$^{23}$Department of Physics and Astronomy, Macquarie University, NSW 2109, Australia \\
$^{24}$University of Alabama, Department of Physics and Astronomy, Tuscaloosa, AL 35487, USA\\
$^{25}$Heidelberg Institute for Theoretical Studies gGmbH, Astroinformatics, Heidelberg, Germany\\
}

\begin{document}

\date{Accepted 2015 July 23.  Received 2015 July 22; in original form 2015 February 27}

\pagerange{\pageref{firstpage}--\pageref{lastpage}} \pubyear{2015}

\maketitle

\label{firstpage}
\clearpage

\begin{abstract}
We present results from the first twelve months of operation of Radio Galaxy Zoo, which upon completion will enable visual inspection of over 170,000 radio sources to determine the host galaxy of the radio emission and the radio morphology. Radio Galaxy Zoo uses $1.4\,$GHz radio images from both the Faint Images of the Radio Sky at Twenty Centimeters (FIRST) and the Australia Telescope Large Area Survey (ATLAS) in combination with mid-infrared images at $3.4\,\mu$m from the {\it Wide-field Infrared Survey Explorer} (WISE) and at $3.6\,\mu$m from the {\it Spitzer Space Telescope}. We present the early analysis of the WISE mid-infrared colours of the host galaxies. For images in which there is $>\,75\%$ consensus among the Radio Galaxy Zoo cross-identifications, the project participants are as effective as the science experts at identifying the host galaxies. The majority of the identified host galaxies reside in the mid-infrared colour space dominated by elliptical galaxies, quasi-stellar objects (QSOs), and luminous infrared radio galaxies (LIRGs). We also find a distinct population of Radio Galaxy Zoo host galaxies residing in a redder mid-infrared colour space consisting of star-forming galaxies and/or dust-enhanced non star-forming galaxies consistent with a scenario of merger-driven active galactic nuclei (AGN) formation. The completion of the full Radio Galaxy Zoo project will measure the relative populations of these hosts as a function of radio morphology and power while providing an avenue for the identification of rare and extreme radio structures. Currently, we are investigating candidates for radio galaxies with extreme morphologies, such as giant radio galaxies, late-type host galaxies with extended radio emission, and hybrid morphology radio sources.
\end{abstract}
\section{Introduction}\label{sec:intro}
Large radio continuum surveys over the past 60 years have played a key role in our understanding of the evolution of galaxies across cosmic time.  These surveys are typically limited to flux densities of $S_{\rm 1.4} \ge 1\,$mJy at $1.4\,$GHz (21~cm), and are consequently dominated by active galactic nuclei (AGN) with $1.4\,$GHz luminosities of $L_{\rm1.4} \ge 10^{23}\,$W~Hz$^{-1}$ \citep[e.g.\ ][]{Mauch2007,Mao12}. The largest such survey, the NRAO VLA Sky Survey \citep[NVSS;][]{Condon1998} is relatively shallow with a completeness level of 50\% at 2.5 mJy~beam$^{-1}$ and 99\% at 3.4 mJy~beam$^{-1}$. For surveys sensitive to flux densities below 1 mJy, the radio emission is a combination of: (1) low-luminosity AGN \citep[$L_{\rm{1.4}} < 10^{22}$ W~Hz$^{-1}$; e.g.\ ][]{Slee94}; and (2) star formation \citep[e.g.\ ][]{condon12}.  Current deep ($S_{\rm 1.4} < 15\,\mu$Jy beam$^{-1}$ ) radio continuum surveys are limited to $< 10$ square degrees of the sky \citep[e.g.,][]{Owen2008,Smolcic2009,condon12,Franzen15} resulting from available observing time.  

Over the next 5 to 10 years, the next generation radio telescopes and telescope upgrades such as  the Australian SKA Pathfinder \citep[ASKAP;][]{Johnston2007}, MeerKAT \citep{Jonas2009} and Apertif \citep{Verheijen08} will perform surveys with higher angular resolution and sensitivity that cover wider fields.  In particular, the wide-area surveys such as the Evolutionary Map of the Universe survey \citep[EMU;][]{Norris2011} using ASKAP; and the WODAN survey \citep{Rottgering11} using the Apertif upgrade on the Westerbork Synthesis Radio Telescope (WSRT) will together provide all-sky coverage to a rms sensitivity of $\approx 10-20$ $\mu$Jy beam$^{-1}$ and cover a large spectral range at $<15$~arcsec resolution.  The combination of EMU and WODAN is expected to detect over 100~million radio sources, compared to the total of $\approx 2.5$ million radio sources currently known.

These widefield surveys will be complemented by deeper field studies over smaller sky areas from facilities such as MeerKAT in the Southern Hemisphere.  Currently,  it is planned that the MeerKAT MIGHTEE survey \citep[e.g.\ ][]{Jarvis12} will reach $\approx1.0$ $\mu$Jy beam$^{-1}$ rms over 35 square degrees of the best-studied extragalactic deep fields accessible from South Africa. Together, these surveys will provide an unprecedented view of activity in the Universe addressing many key science questions on the evolution of AGN and star formation in galaxies as well as the cosmic large-scale structure.

To harvest new scientific knowledge from these very large surveys, the detected radio sources need to be cross-identified with galaxies observed at other wavelengths.  The task of cross-matching a radio source with its host galaxy is complicated by the large and complex radio source structures that are often found in radio-loud AGN.  For survey samples of several thousand sources, radio cross-identifications have traditionally been performed through visual inspections \citep[e.g.][]{Norris06,middelberg08,Gendre2010,Lin10}.  Automated radio classification algorithms are still in the infancy stage; \citet{Norris2011} estimated that approximately $10\%$ of the 70 million radio sources expected from the EMU survey will be too complicated for current automated algorithms \citep[e.g.\ ][]{Proctor06,Kimball2008,vanvelzen15,Fan15}. Importantly, these complex sources are also likely to be those with the greatest scientific potential.

To test possible solutions to this cross-identification issue, we have created Radio Galaxy Zoo\footnote{\tt{http://radio.galaxyzoo.org}}, an online citizen science project based upon the concepts of the original Galaxy Zoo \citep{Lintott2008}.  Following its launch in 2007, the success of the Galaxy Zoo project inspired the creation of the Zooniverse\footnote{\tt{http://www.zooniverse.org}}, now a highly successful platform for online citizen science, hosting more than 30 projects across a diverse selection of research areas (from astronomy, to history and biology), and with over 1.4 million users. Zooniverse projects share a common philosophy of ``real research online'' with a clear research goal and a real need for human input. The first project, Galaxy Zoo, has produced over 50 peer reviewed publications to date \citep[for a recent summary, see e.g.\ ][]{Fortson12}. 

In Radio Galaxy Zoo, the public is asked to cross-match radio sources, often with complex structures, to their corresponding host galaxies observed in infrared images. The importance and complexities of radio source morphologies are described in Section~\ref{sec:morph}. Section~\ref{sec:rgz} describes the Radio Galaxy Zoo project.  Early analyses of the reliability of Radio Galaxy Zoo source cross-identifications and classifications are discussed in Section~\ref{sec:analysis}.  Section~\ref{sec:science} presents the science outcomes obtained from the first year of project operation. We summarise our project and early results in Section~\ref{sec:summary}. Throughout this paper we adopt a $\Lambda$CDM cosmology of $\Omega_m = 0.3$, $\Omega_{\Lambda}=0.7$ with a Hubble constant of $H_{\rm{0}}=70$~km~s$^{-1}$~Mpc$^{-1}$.
\section{Radio source morphologies}\label{sec:morph}
In low-redshift radio sources, a combination of radio morphology and radio spectral index is useful for distinguishing whether the observed radio emission is dominated by star formation or AGN with the presence of core-jet, double- or triple- radio sources providing evidence for AGN-dominated emission. On the other hand,  the combination of radio and infrared observations prove to be the most effective means for differentiating between AGN- and star formation-dominated emission at higher redshifts \citep[e.g.\ ][]{Seymour2008,Seymour2009}.

A key step in determining radio source physical properties is the determination of their distance through redshifts associated with the identification of the host galaxies from which the radio emission originate. The difficulty in cross-identification can be exemplified by the case of a linear alignment of three radio sources \citep[e.g.\ ][]{Norris06} which can be either: (1) a chance alignment of radio emission from three separate galaxies; (2) three radio components from a single radio-loud AGN with two extended radio lobes; or (3) the chance alignment of a double radio source and a compact radio source.

While the vast majority ($\sim90\%$) of radio sources are compact in structure \citep{Shabala08,Sadler14}, the extended morphologies of radio-loud sources were first classified by \citet{Fanaroff1974} based on 57 sources from the Third Cambridge (3C) Radio sample \citep{Mackay1971}.  Fanaroff \& Riley (1974) separated their sample of radio galaxies according to the ratio of the distance between the regions of highest brightness on opposite sides of the host, to the total source extent from one end to the opposite; a ratio below 0.5 was class I, and a ratio above 0.5 was class II, now known as the ``Fanaroff--Riley'' types \FRI\, and \FRII, respectively.  They also found a sharp division in radio luminosity density between the two classes at $L_{178\,{\rm MHz}} \approx 2 \times 10^{25}\,$W Hz$^{-1}$ sr$^{-1}$, with \FRII\, sources above and \FRI\, sources below this luminosity density.   This classification was later confirmed by \citet{Owen1994} who found that this break between \FRII\, and \FRI\, radio sources also correlates with optical luminosity. However, this correlation with the optical luminosity consists of significant overlap between the two populations \citep[e.g.\ ][]{best09}.  Further investigation into \FRII\, and \FRI\, sources has produced a number of different radio source morphologies. Unusual classes of radio source morphologies include Narrow Angle Tail \citep[NAT; ][]{Rudnick1976}, Wide Angle Tail \citep[WAT; ][]{Owen1976}, and Hybrid Morphology Radio Sources \citep[HyMoRS;][]{Gopal2000}.  The NAT sources are usually thought to have high peculiar velocities \citep[e.g.\ ][]{Venkatesan94}, whereas WATs are mostly associated with galaxy clusters where the ICM density and the relative motions of the cluster galaxies are responsible for the shape and structure of the observed radio sources \citep[e.g.\ ][]{Owen1976,Rudnick1976,Burns98}. The current method of determining the morphology of extended radio sources is by visual inspection and as a result this is only applicable to samples of no more than a few thousand radio sources \citep[e.g.\ ][]{middelberg08}.

\begin{center}
\begin{figure*}
\includegraphics[scale=0.55]{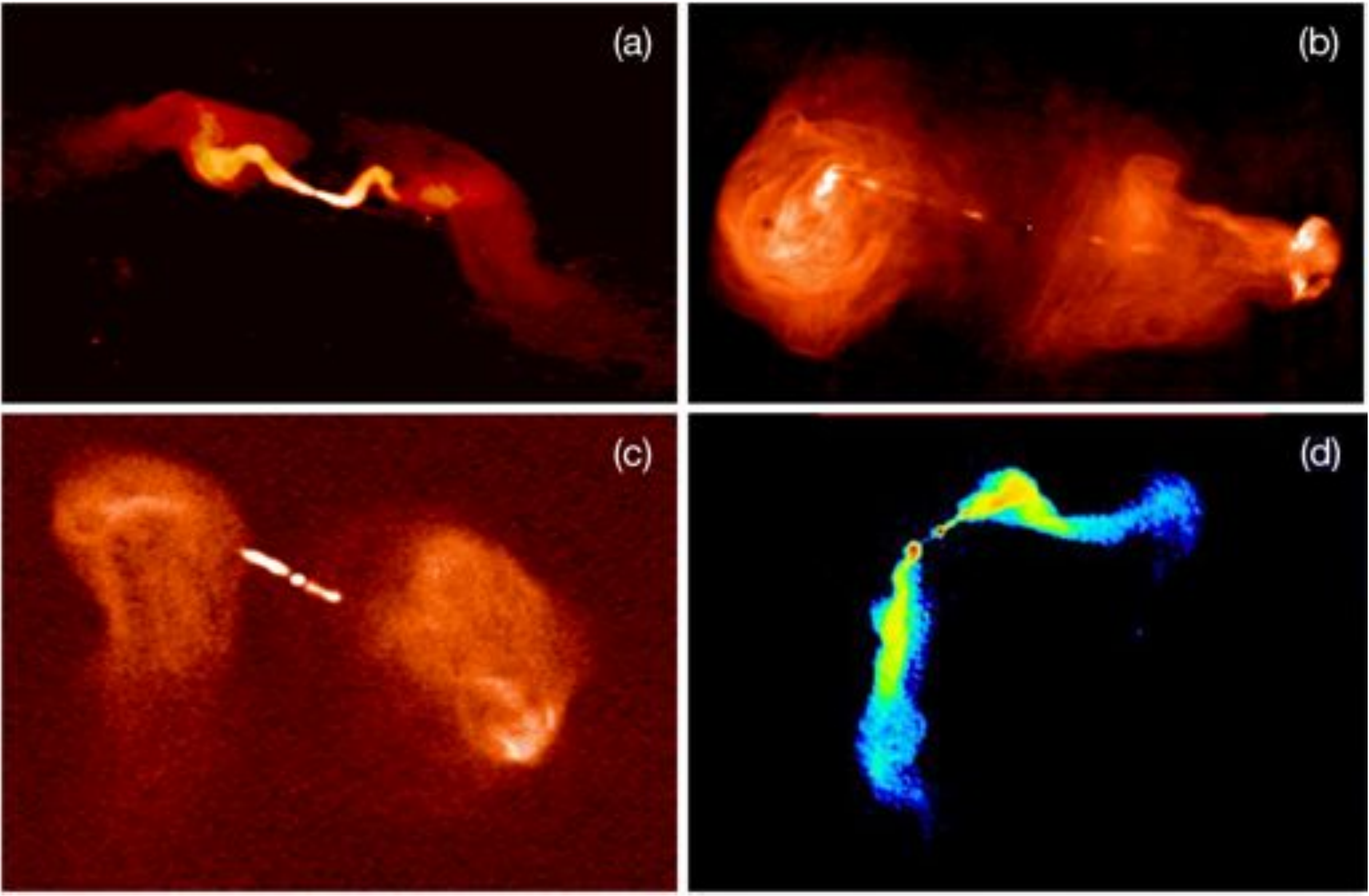}
\caption{Four examples of various radio loud galaxy morphologies.  (a) \FRI\, radio source 3C31 at $1.4\,$GHz with the VLA from NRAO/AUI (http://images.nrao.edu/AGN/Radio\_Galaxies/) by R.~Laing, A.~Bridle, R.~Perley, L.~Feretti, G.~Giovannini, and P.~Parma \citep{Laing96}.  (b) 3C353 at $8.4\,$GHz with the VLA from NRAO/AUI by M.~Swain, A.~Bridle, and S.~Baum \citep{Swain1998}.  (c) 3C288 at $8.4\,$GHz with the VLA from NRAO/AUI by A.~Bridle, J.~Callcut, and E.~Fomalont \citep{Bridle89}.  (d) A WAT radio source, 3C465, in Abell 2634 at $1.4\,$GHz with the VLA from the Atlas of DRAGNs (http://www.jb.man.ac.uk/atlas/object/3C465.html) by F.~Owen \citep{Eilek84,Leahy1996}.}
\label{fig:morph}
\end{figure*}
\end{center}

 In  Fig.~\ref{fig:morph} we present four examples of extended radio morphologies that can be found in galaxies. 
Fig.~\ref{fig:morph}(a) shows an example of a \FRI\, radio source, 3C31 from NRAO/AUI by R.~Laing, A.~Bridle, R.~Rearly, L.~Feretti, G.~Giovannini, and P.~Parma. Fig.~\ref{fig:morph}(B) shows a \FRII\ radio source, 3C353, with hotspots in both radio lobes as well as the narrow jet and counterjet from NRAO/AUI by M.~Swain, A.~Bridle, and S.~Baum.  We show radio source, 3C288, with more complex structures in Fig.~\ref{fig:morph}(c) from NRAO/AUI by A.~Bridle, J.Callcut and E.~Fomalont. 3C288 exhibits an unusual asymmetry in its radio morphology and edge-darkening can only be observed on one side of this double-lobed radio source.  Finally, Fig.~\ref{fig:morph}(d) shows 3C465, an example of a WAT source from the Atlas of DRAGNs \citep{Leahy1996} by F.~Owen.

\section{Radio Galaxy Zoo}\label{sec:rgz}
Radio Galaxy Zoo is an online citizen science project where volunteers classify radio galaxies and their host galaxies via a web interface.  The main purpose of Radio Galaxy Zoo is to produce cross-identifications for resolved radio sources which are too complex (i.e.\ where the two radio lobes are widely separated or where the radio morphology is asymmetrical or otherwise complex) for automated source matching algorithms \citep[e.g.\ ][]{Becker95,Mcmahon02,Kimball2008,Proctor11,vanvelzen15}. In the current phase of the project, we are offering to the volunteers a total of 177,218 radio sources from two radio surveys described in the following subsections.  To address our need for the classifications of complex radio source morphologies, we have biased our sample against unresolved sources as described in Section \ref{sec:first data}.

Although initially designed as a pilot study in preparation for the 7~million complex radio sources from the upcoming EMU survey, we are currently exploring the inclusion of other radio surveys for subsequent phases of this project.  In addition to being an alternative technique to processing large datasets, the result of Radio Galaxy Zoo will also provide an ideal training dataset for the development and implementation of future-generation machine-learning algorithms in the field of pattern recognition.

\subsection{Data}
We extracted the radio sources for this project from the Faint Images of the Radio Sky at Twenty Centimeters \citep[FIRST;][]{White1997,Becker95} and the Australia Telescope Large Area Survey Data Release 3 (ATLAS; Franzen et al. (2015) submitted to MNRAS). We chose FIRST over NVSS \citep{Condon1998} due to its higher resolution and greater depth, making it more comparable to ATLAS and EMU. We expect many of these radio sources to be at high redshifts so observations of the host galaxies' stellar components are typically derived from infrared surveys to reduce the effects of dust obscuration.  In our case, we offer overlays of the FIRST and ATLAS fields to equivalent fields in the mid-infrared wavelengths from the {\it Wide-Field Infrared Survey Explorer} \citep[WISE;][]{Wright2010}  and the {\it Spitzer} Wide-Area Infrared Extragalactic Survey \citep[SWIRE;][]{Lonsdale2003} surveys, respectively.

\subsubsection{FIRST and WISE data}\label{sec:first data}
The majority of the data in Radio Galaxy Zoo comes from the $1.4\,$GHz FIRST survey (catalogue version 14 March 2004) and the $3.4\,\mu$m WISE survey \citep[all-sky data release in March 2012; ][]{Cutri13}.  FIRST covers over 9000 square degrees of the northern sky down to a $1\sigma$ noise level of $150\,\mu$Jy beam$^{-1}$ at $5\,\arcsec$ resolution.  WISE is an all-sky survey at wavelengths $3.4$, $4.6$, $12$, and $22\,\mu$m with $5\sigma$ point source sensitivity in unconfused regions of no worse than $0.08$, $0.11$, $1.0$, and $6.0\,$mJy \citep{Wright2010}.  These four wavebands are also identified as $W1$, $W2$, $W3$ and $W4$ in order of increasing wavelength.  The selection of these four bands makes WISE an excellent instrument for studies of stellar structure and interstellar processes of galaxies.  The two shorter bands trace the stellar mass distribution in galaxies and the longer wavelengths map the warm dust emission and polycyclic aromatic hydrocarbon (PAH) emission, both tracing the current star formation activity.

\begin{center}
\begin{figure}
\includegraphics[scale=0.38]{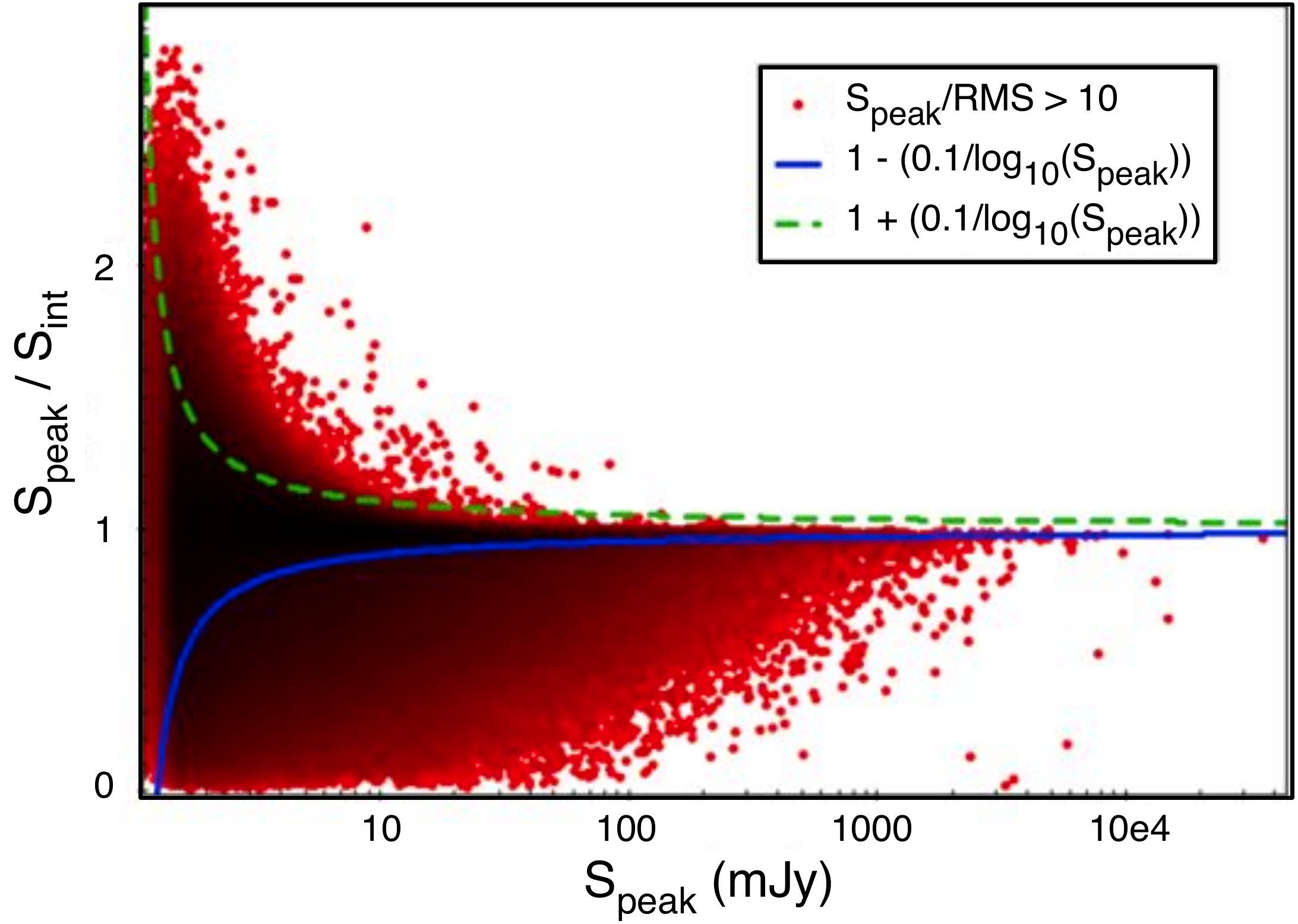}
\caption{The distribution of peak to integrated flux density ratio as a function of peak flux density for all FIRST radio sources with SNR$>10$. The scatter in the flux ratio above the $S_{\rm{peak}}/S_{\rm{int}}=1$ line at low fluxes is the result of intrinsic errors on the peak and total flux density measurements. The points below the solid blue line represents the Radio Galaxy Zoo selection of extended sources.  The mirror of this line around $S_{\rm{peak}}/S_{\rm{int}}=1$ is shown by the green dashed line and demonstrates that a fraction of the selected sources will be compact. We estimate that our sample contains approximately 16\% compact sources for control purposes.} 
\label{fig:select}
\end{figure}
\end{center}

We designed Radio Galaxy Zoo to cross-match complex radio sources with their host galaxy rather than simple, compact radio sources which are easily matched by algorithms. We filtered the FIRST radio catalogue based on two criteria: (1) the radio source has a signal-to-noise ratio (SNR) greater than 10; and (2) the radio source is considered to be resolved. We considered a source to be resolved if it satisfies the criterion:
\begin{equation}
\label{eqn:first}
\frac{S_{\rm{peak}}}{S_{\rm{int}}} < 1.0 - \left(\frac{0.1}{{\rm log}(S_{\rm{peak}})}\right) \, ,
\end{equation}
\noindent where $S_{\rm{peak}}$ is the peak flux density in mJy~beam$^{-1}$ and $S_{\rm{int}}$ is the total flux density of the radio source in mJy.  This selection criterion is indicated by the blue solid line in Fig.~\ref{fig:select} and selects 218,228~radio sources from the FIRST catalogue. At low peak flux densities the scatter around $S_{\rm{peak}}/S_{\rm{int}} = 1$ rapidly increases due to intrinsic measurement errors on the peak and total fluxes, e.g. leading to unphysical situations where $S_{\rm{peak}} > S_{\rm{int}}$. The larger number of sources below the $S_{\rm{peak}}/S_{\rm{int}} = 1$ line corresponds to real extended sources. Assuming that the 34,689 radio sources found in the area that is mirroring the relation (the green dashed line) represents the 34,916 (16\%) compact sources that can be expected in our sample and are useful for control purposes.  At the time of publication, a random subset of 174,821 out of the 218,228 fields of $3\arcmin \times 3\arcmin$ from the FIRST survey have been made available to Radio Galaxy Zoo participants.

\subsubsection{ATLAS and SWIRE}
The 4396 radio sources drawn from ATLAS cover 6.3 square degrees with $2.7\,$square degrees centred on the European Large Area {\it ISO} Survey South 1 field (ELAIS S1) and 3.6 square degrees centred on the {\it Chandra} Deep Field South (CDFS).  ATLAS reaches a $1\sigma$ noise level of $16\,\mu$Jy beam$^{-1}$ in ELAIS S1 and $13\,\mu$Jy beam$^{-1}$ in CDFS \citep{Franzen15}. The angular resolution of the survey varies across the two regions with a mean of $12.2\arcsec \times 7.6\arcsec$ in ELAIS S1 and $16.8\arcsec \times 6.9\arcsec$ in CDFS.  ATLAS was chosen because the two fields are considered the pilot fields for the EMU survey and as such the resolution and sensitivity limits are comparable to EMU.  The $3.6\,\mu$m images come from the SWIRE survey which covers 6.58 square degrees centred on CDFS and 14.26 square degrees centred on ELAIS S1 at 3.6, 4.5, 5.8, and $8.0\,\mu$m down to a $5\sigma$ noise level of 7.3, 9.7, 27.5, and $32.5\,\mu$Jy \citep{Lonsdale2003}. A random subset of 2,397 radio sources from ATLAS are currently offered to Radio Galaxy Zoo's participants. 

\begin{center}
\begin{figure}
\includegraphics[scale=0.7]{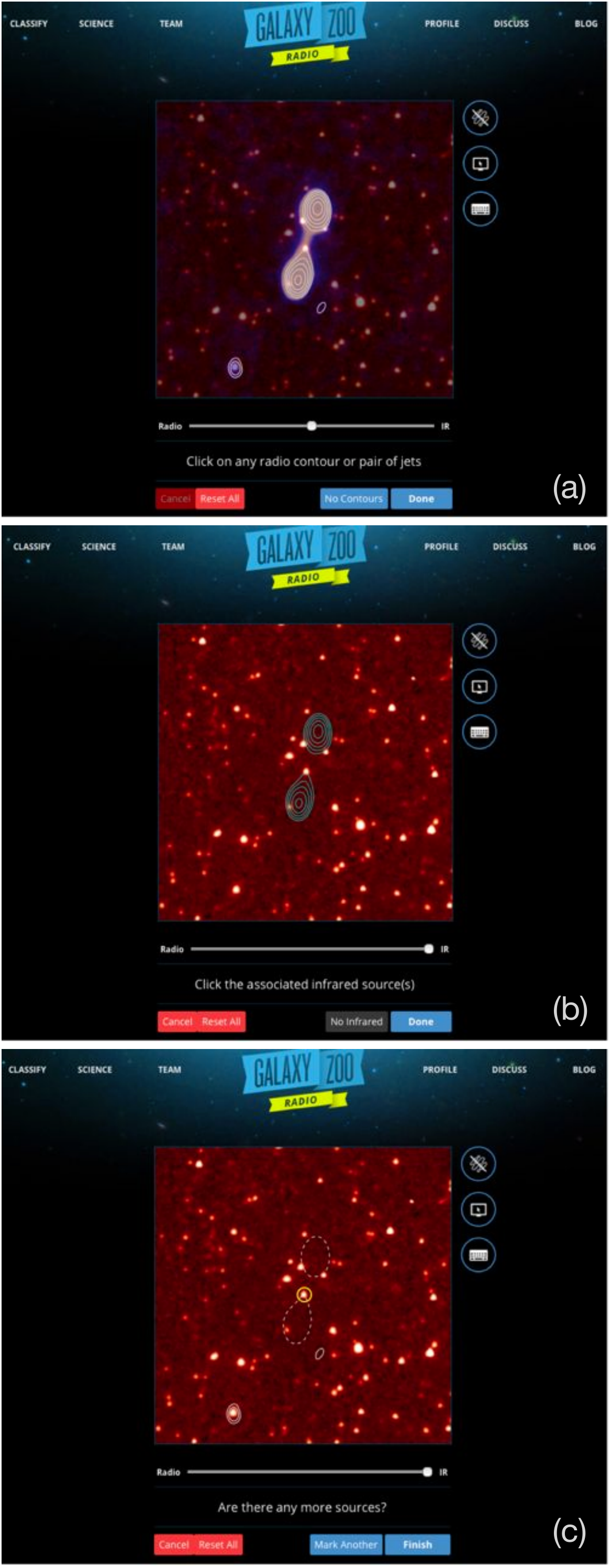}
\caption{The Radio Galaxy Zoo interface illustrating the three steps required to make a classification. A single $3\arcmin \times 3\arcmin$ field-of-view, which is designated as the ``subject''. (a) Step 1: select the radio components that belong to a single radio source.  (b) Step 2: select the associated infrared galaxy that corresponds to the selected radio source.  (c) Step 3: either continue classifying the remaining radio sources in the image or move on to the next subject. All images were obtained from http://radio.galaxyzoo.org.} 
\label{fig:inter}
\end{figure}
\end{center}

\subsection{Interface description}

Radio Galaxy Zoo was launched on December 17th 2013. This international online citizen science project is available in 8 languages (English, Spanish, Russian, Chinese, Polish, French, German and Hungarian) and invites participants to match radio sources with the corresponding infrared host galaxy following a decision tree similar to the original Galaxy Zoo project \citep{Lintott2008}. While Galaxy Zoo uses colour composite images of Sloan Digital Sky Survey (SDSS), Radio Galaxy Zoo enables the participant to transition between the mid-infrared image and the radio 1.4~GHz image via a slider.

The Radio Galaxy Zoo interface is shown in Fig.~\ref{fig:inter}.  The radio and infrared images are overlaid upon one another with the lowest contour and shading for the radio images pre-set at a $3\sigma$ level, as shown by the blue contours in Fig.~\ref{fig:inter}(a) and (b).   There is a continuum of transparency levels between the radio and infrared images with the default transparency position in the middle.  When a participant transitions from the radio to the infrared image, the radio colour map will be gradually replaced by a set of contours (as shown in Fig.~\ref{fig:inter}b). The interface also includes a spotter's guide containing examples of radio sources matched to infrared sources, keyboard shortcuts, a toggle function to turn on or off the radio contours, and a link to return to the tutorial.  

At the beginning, each participant is introduced to the project through the completion of a simple tutorial which guides them through the necessary steps to complete the classification of a single subject.  The participant is required to follow three steps to make a classification: (1) select the radio contours that the participant considers to correspond to one radio source (Fig.~\ref{fig:inter}a); (2) select the corresponding infrared host galaxy which corresponds to the selected radio contours (Fig.~\ref{fig:inter}b); and (3) either continue classifying the remaining radio sources or progress to the next image (Fig.~\ref{fig:inter}c).  For each step, the tutorial contains information on how to select the correct part of the image.  After completion of the tutorial a randomly selected image from the Radio Galaxy Zoo data set is immediately selected so that participants begin working on real data as soon as possible.

Each Radio Galaxy Zoo subject is only offered once to each participant and is subsequently withdrawn from being offered once the subject reaches a given threshold of classifications;  this threshold is dependent on the complexity of the source. For sources with a single and/or connected set of radio contours, the vast majority of sources are expected to have a single IR galaxy counterpart (with the exception of blended sources that may originate from separate host galaxies).  We record the nearest IR source to the participants' clicks as the host galaxy.  Such an identification requires fewer independent classifications, and so these images are retired from the interface after 5~classifications. For the remainder of the images, which have multiple radio components, a higher threshold of 20~classifications is adopted for higher accuracy.  The data is stored in a {\tt mongoDB} database structure with each click on the image recorded for each step. We record the positions of the corners of the box surrounding the selected radio contours and the position of the selected infrared host galaxy. 

After completion of the classification and prior to progressing to the next radio source, particularly engaged participants can opt to discuss the subject in further detail through the RadioTalk forum.  RadioTalk includes links to larger (9\arcmin $\times$ 9\arcmin) FIRST and WISE images, images from NVSS \citep{Condon1998} and optical observations from SDSS Data Release 10 \citep{SDSS10} and SDSS Data Release 12 \citep{Alam15} for further detailed investigation. There is also a discussion board used for discussions on an object and for general help on the project as a whole.  This is where the interaction between the science team and the volunteers occurs and many new candidate discoveries are further investigated. In Galaxy Zoo, these forum discussions resulted in the discovery of new classes of objects such as the Voorwerpjes and ``Hanny's Voorwerp'' -- an ionization light echo from a faded AGN \citep{Lintott2009,Keel2012} as well as the ``green peas''-- [OIII] emission line-dominated compact star-forming galaxies \citep{Cardamone09}. 

\subsection{User Base}
\begin{center}
\begin{figure}
\includegraphics[scale=0.42]{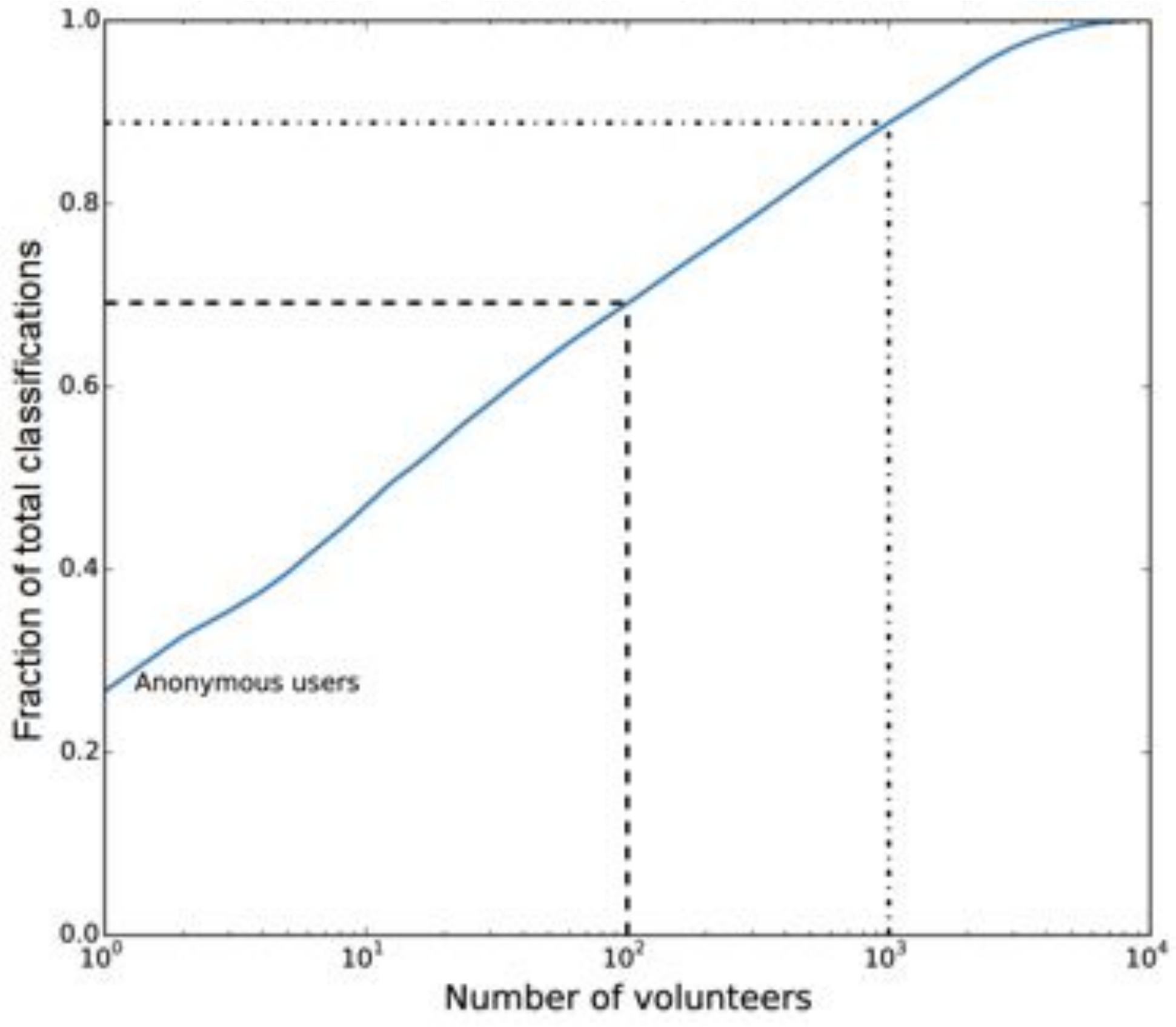}
\caption{Cumulative distribution of the total number of classifications in Radio Galaxy Zoo as of 1 May 2015. Anonymous users who have not logged into the interface are responsible for 27\% of classifications; the top 100 registered users (dashed lines) have done an additional 42\% of the total, while the top 1,000 users (dot-dashed lines) are responsible for 62\% of the registered classifications.}
\label{fig:users}
\end{figure}
\end{center}

On May 1, 2015, Radio Galaxy Zoo had over 6900 registered volunteers and 1,155,000 classifications.  Each participant has the option of logging into the Zooniverse system which benefits the Radio Galaxy Zoo project by allowing us to identify the contributions made by individuals.  There are 102~participants (1.4\%), each of whom has classified over 1,000 subjects, and 11 of these (0.15\%) who have classified over 10,000 subjects.  Fig.~\ref{fig:users} shows the distribution of classifications of participants in the project. More than half (62\%) of our project is completed by the top 1,000 volunteers (in terms of the number of subjects classified). Participants who choose not to log into the system still have their classifications recorded; in the absence of other information, we use their IP addresses as substitute IDs. Anonymous users have generated 26.8\% of the total classifications to date. 

\section{Early Data Analysis}\label{sec:analysis}
\subsection{Control Sample}\label{sec:controlsample}
\begin{figure*}
\includegraphics[scale=0.85]{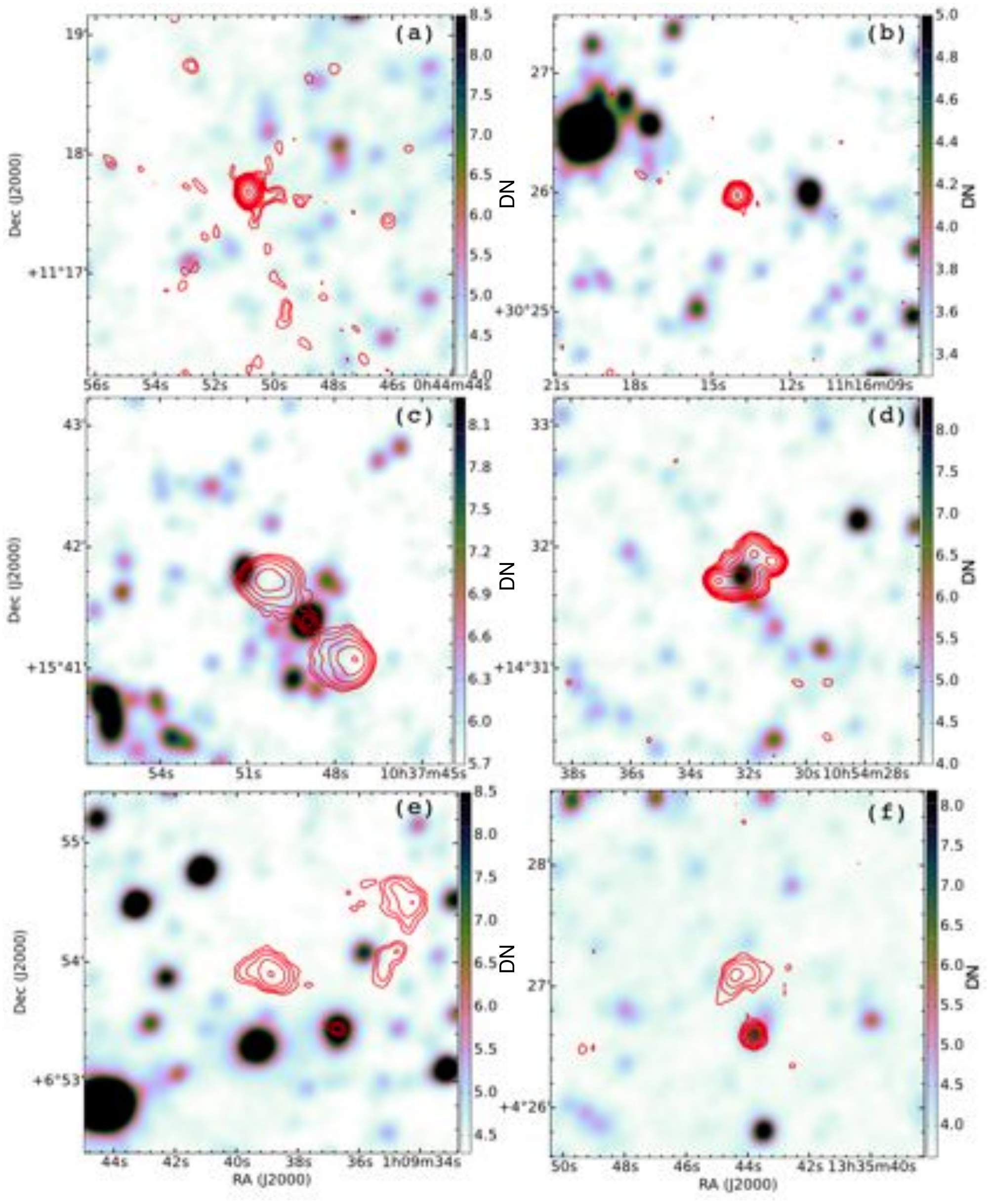}
\caption{Example images of the different types of radio sources in our control sample (as described in Section \ref{sec:controlsample}).  (a) A compact source with image processing artefacts. We included a number of compact radio sources in our control sample for the purpose of a consistency check.  (b) Compact radio source.  (c) Double-lobed radio source.  (d) Bent morphology radio source.  (e) A wide angle tail.  (f) Unusual one-sided core--lobe radio source.  The background image in all panels is the WISE 3.6~$\mu$m image and the contours are the FIRST image started at 3 times the local rms and increasing in multiples of 2.  Each image is $3\arcmin \times 3\arcmin$ in size.  The background colour scheme comes from {\tt CUBEHELIX} \citep{Green2011}.}
\label{fig:expert}
\end{figure*}

To perform a preliminary assessment of the morphological classifications currently completed by the Radio Galaxy Zoo volunteers, we use a collection of 100~images  classified by 10~members of the science team (Banfield, Kapi\'{n}ska, Masters, Middelberg, Rudnick, Schawinski, Shabala, Simmons, Willett \& Wong) as our control sample. Of the 100 subjects, 57 are selected to represent a range of complex and/or unusual morphologies seen in radio galaxies, including double and triple sources, bent and precessing jets, HyMoRS, and artifacts (see Fig.~\ref{fig:expert}). The remainder consisted of randomly selected images already classified by at least 20~volunteers; many of these include emission from compact radio morphologies with a single component. 

Our control sample of 100~subjects serves two purposes. The first is to compare the levels of agreement among the experts, which is critical for establishing cutoffs in the development of consensus algorithms. We establish the limit on the consensus by identifying the classes of Radio Galaxy Zoo subjects that are too complicated for a consensus to be reached by the expert science team. Classifications are separated into three categories corresponding to the vote fraction and consensus level of the classifiers:
\begin{itemize}
    \item Class A: all or all but one expert classifier(s) agree on the number of radio components per radio source and the location(s) of the IR counterpart;
    \item Class B: 2  experts disagreed on the number of radio components or IR counterparts; and
    \item Class C: 3 or more experts did not agree on the fundamental radio/IR morphology.
\end{itemize}

\noindent Of the 100~images, the science team classifications had 53 in Class A, 31 in Class B, and 16 in Class C. Individual inspection of the classifications for the Class C images reduced them to a final 10 that the science team agreed would require genuine follow-up observations to distinguish between morphological categories. We assume that both Classes A and B meet thresholds for a unique classification, which we verify using joint inspection by the entire team. We thus tentatively assume that 90 per cent accuracy is the highest possible level that can be expected from group classification either from volunteers or experts.  The cutoff for consensus on individual subjects can vary depending on the number of radio components and relative difficulty of the classification.

\begin{figure}
\includegraphics[scale=.33]{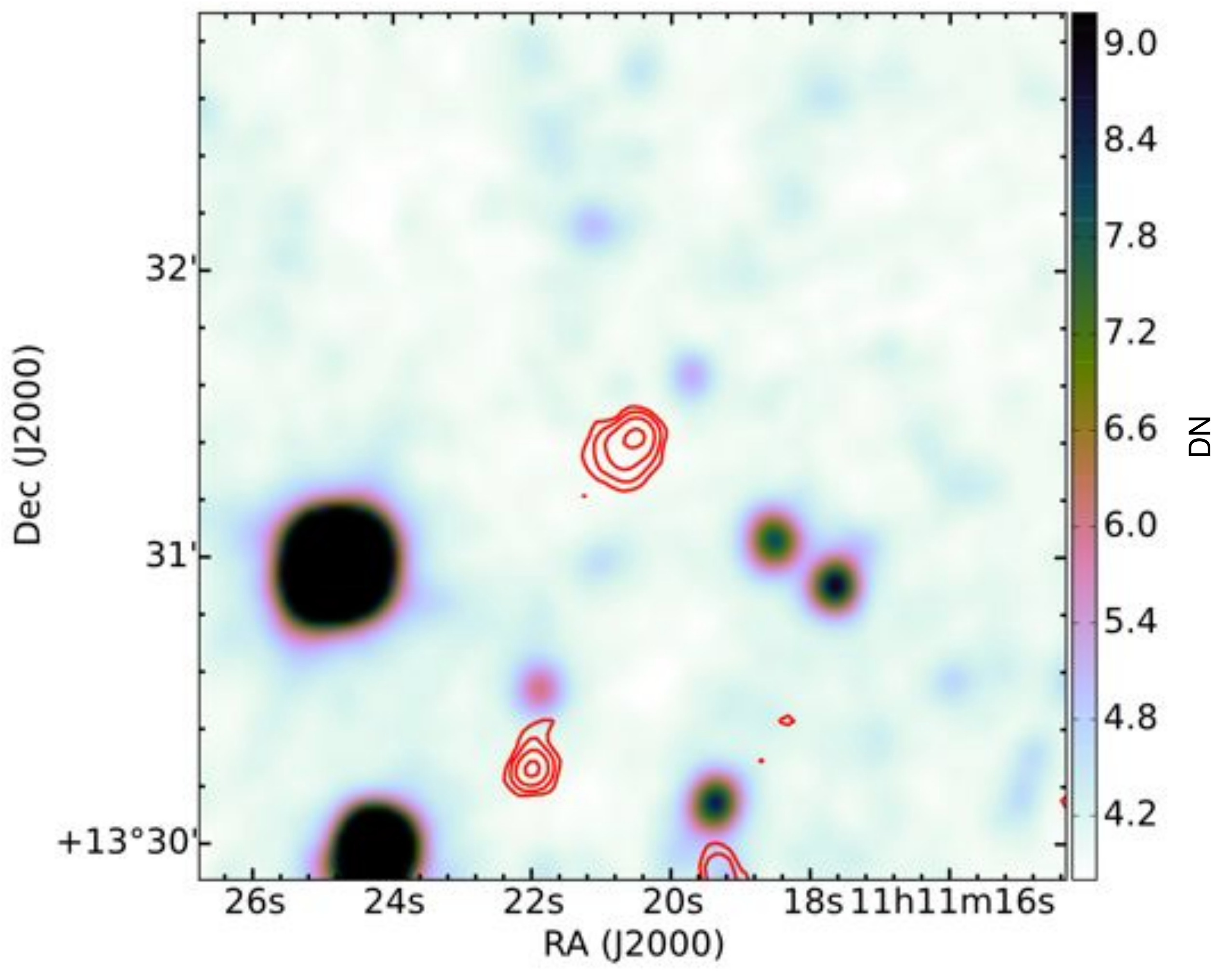}
\caption{FIRSTJ111120.5+133123 --- an example where there was no agreement between the expert panel on the source identification. }
\label{fig:noconsensus}
\end{figure}

Fig.~\ref{fig:noconsensus} shows an example where the expert members of the science team could not reach a consensus on whether this subject contains two independent sources or a single double source, and whether the IR host was visible.  In cases where there is significant disagreement between experts or volunteers, these sources will be deferred to further study where Bayesian-type analyses will assign a probability that the host has been correctly identified.  These Bayesian analyses will be based on: (1) the probability that the two radio sources near the center are in fact part of a double; (2) the separation of the host from the centroid of the radio emission; and (3) the luminosity and the colours of the host. 

\begin{figure}
\includegraphics[scale=.33]{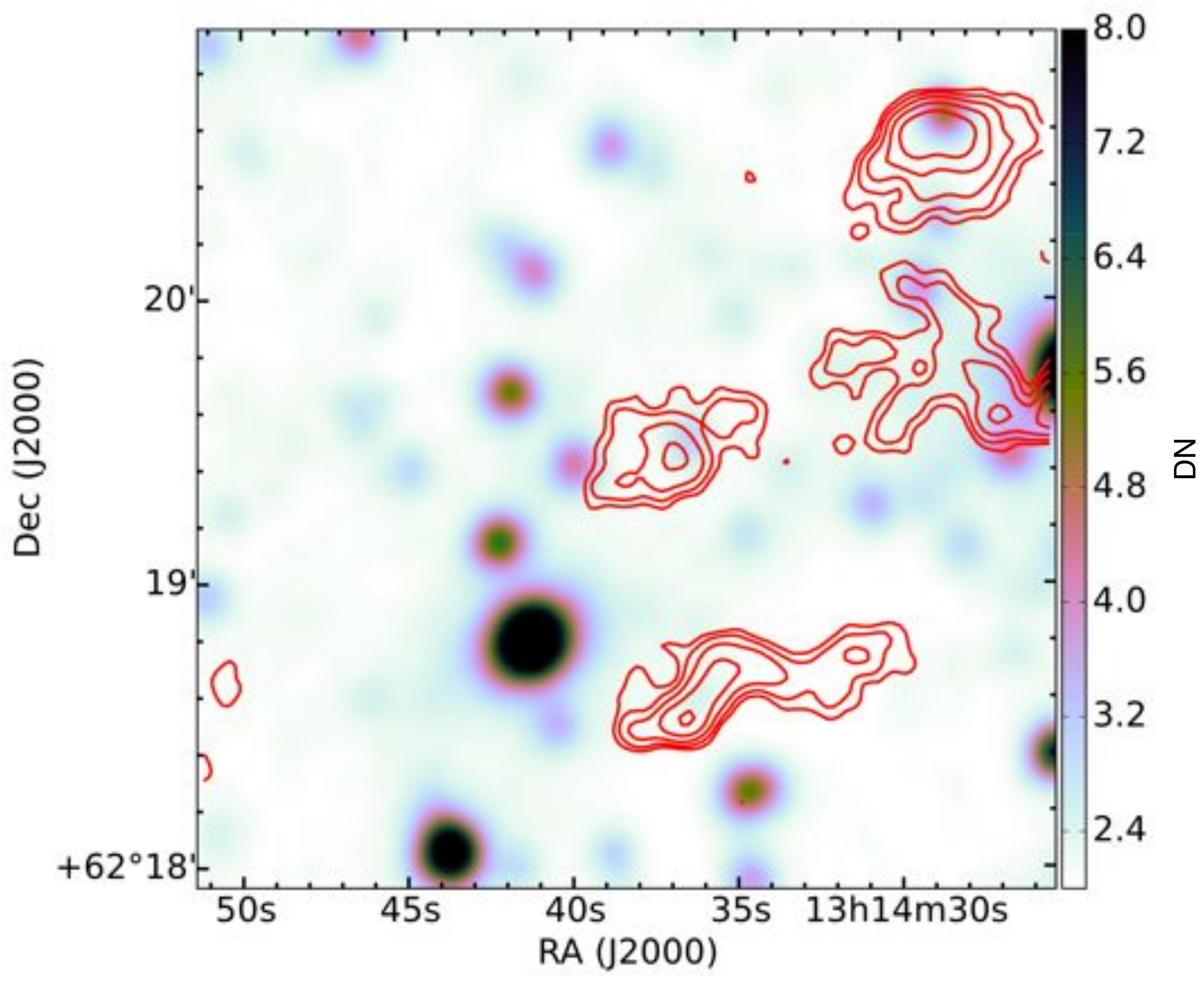}
\caption{An example of an unusual radio morphology identified by the Radio Galaxy Zoo volunteers. It is unclear whether all the radio sources in this subject are discrete components of the same source or if they are indeed independent sources. In this case, the host galaxy lies beyond the field of view of this subject.}
\label{fig:unexpected3}
\end{figure}

We find that the source identifications from Radio Galaxy Zoo volunteers are as likely to disagree as the experts for difficult or `unusual' radio sources. Fig.~\ref{fig:unexpected3} shows an example field where there are multiple radio components.  Although there is one clear identification with the bright elliptical (SDSS~J131424.68+621945.8) in a cluster of galaxies at $z=0.131$,  it is unclear how many individual sources there are in this image, or whether these are all detached pieces of the same radio galaxy, now being energized by turbulence or shocks in the intracluster medium.

The second goal of the control sample was to assess the accuracy of the volunteers, both by looking at their relative agreement levels and by comparing their results to the expert classifications. We measured the consensus for a subject using $C = n_{\rm consensus}/n_{\rm all}$, where $n_{\rm consensus}$ is the number of volunteers who agreed on the arrangement and host galaxy ID for every radio component in the image, and $n_{\rm all}$ is the total number of classifications for the image. We find that the mean consensus level is $C=0.67$, indicating that the majority of images do have a single majority classification (without necessarily confirming whether this consensus is in fact correct). More than 75 per cent of the images in the control sample had $C>0.50$, where the consensus included a majority of independent classifiers. We also found that the consensus is strongly related to the complexity of the image being classified. When there was only one radio source in the image, the mean consensus was $C=0.73$ whereas for complex images with more than one radio source component, the mean consensus for the volunteers was $C=0.44$. 

\begin{figure}
\includegraphics[scale=.35]{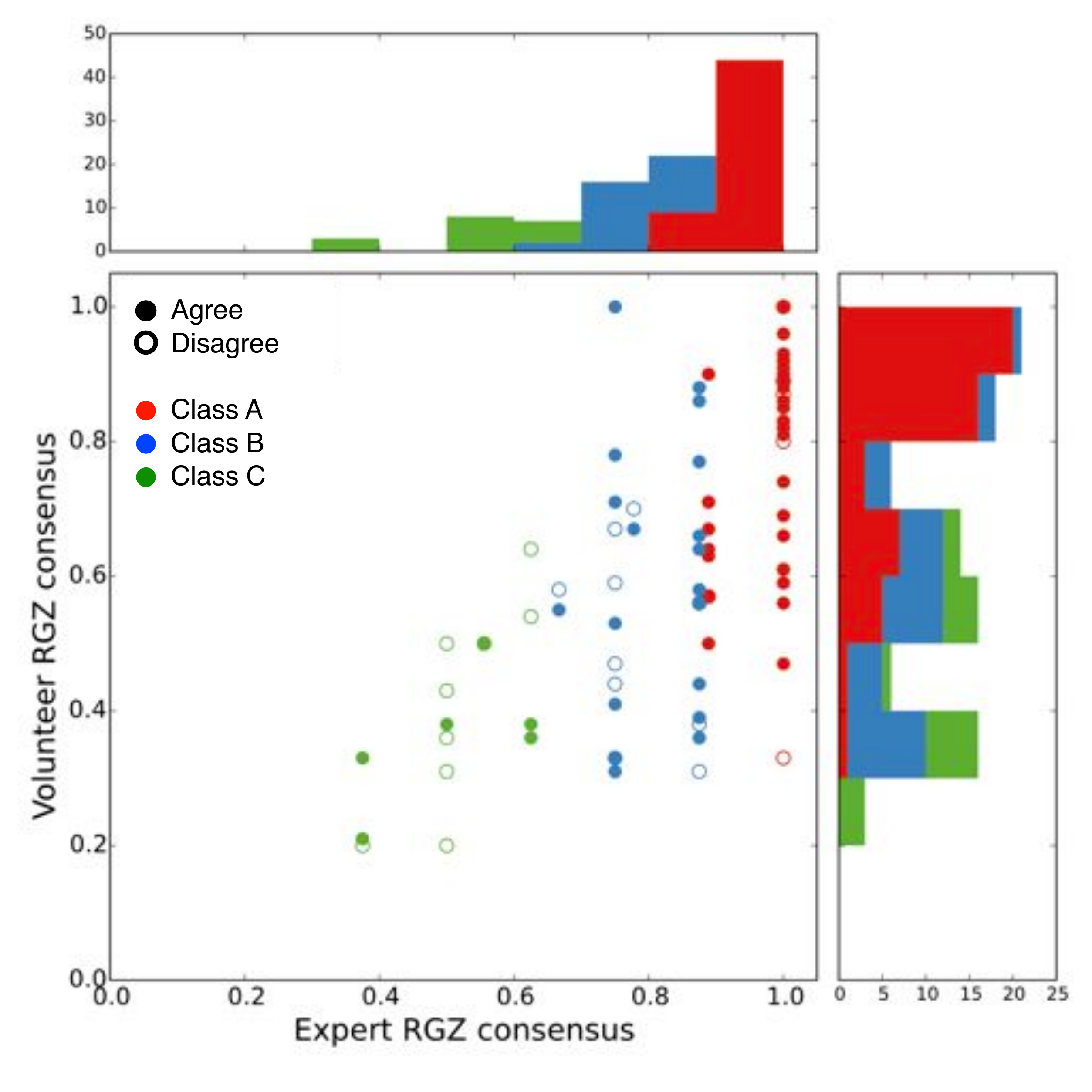}
\caption{Consensus metrics for the 10 Radio Galaxy Zoo experts and volunteers for the control sample of 100 subjects with each point representing one radio source in our control sample. Filled circles show galaxies for which the consensus for both experts and volunteers exactly matched; open circles indicate if they disagreed in any way. Galaxies characterized as classes A, B, or C by the expert science team (Section~\ref{sec:controlsample}) are plotted in red, blue, and green, respectively.}
\label{fig:scatter}
\end{figure}

While the consensus categories of Class A, B or C provides a confidence level for the classifications made by both the experts and the volunteers, a ``confident'' classification does not necessarily mean that the specific cross-identifications made by both the experts and the volunteers will agree.  Hence, we also measure how well the volunteers agree with the experts for the 100 subject control sample. For 74 of the 100 control images, the consensus vote from the volunteers was the same as that selected by the science team (see Fig.~\ref{fig:scatter}).   We note that the control sample was deliberately selected to have a high percentage of morphologies which are difficult to classify, and so we expect the volunteers' performance on the full sample to significantly exceed this. The agreement of the consensus is also a strong function of the \emph{expert} level of agreement.  For classes A, B, and C, the consensus classification concurs with experts 83 per cent, 50 per cent, and 36 per cent of the time, respectively. Disagreements between volunteers and experts with high levels of consensus are mostly driven by the identification of the IR source, rather than the radio components. For example, 10 of 16 Class A or B subjects (as labelled by experts) with which the volunteers disagreed were due to either mispositioning of the IR counterpart or identification of a low $S/N$ IR peak where the experts identified no source in the image.  Table~\ref{tbl-accuracy100} compares the classification distributions between the experts and the volunteers.  

\begin{table} 
\caption{Classification distributions of experts vs. volunteers for the control sample of 100 subjects in Radio Galaxy Zoo. Experts and volunteers agreed on the plurality classification for 74 out of 100 galaxies; most disagreements were for cases where the experts are in better agreement than the volunteers or where the image has a complicated, Class C morphology. The plurality classification is the classification with the most classifications.  \label{tbl-accuracy100}}
 \begin{tabular}{@{}lcrrr}
 \hline
\hline
\multicolumn{1}{c}{} &
\multicolumn{1}{c}{Volunteers} &
\multicolumn{1}{c}{A} &
\multicolumn{1}{c}{B} &
\multicolumn{1}{c}{C}
\\ 
Experts &     &   &      \\
\hline

Agreed \\
A & &      24      &   14      &       9 \\
B & &       2      &    6      &      13 \\
C & &       0      &    0      &       6 \\
\hline
Disagreed \\
A & &       2      &    2      &       2 \\
B & &       0      &    2      &       8 \\
C & &       0      &    0      &      10 \\ 
\hline
\hline
\end{tabular}
\end{table}

\subsection{Consensus algorithms}\label{sec:consensus}
The underlying data reduction relies on independent classifications by distinct users, and no individual subject is inspected by the same user more than once. The validity of the single classification assumption is straightforward to verify for the 883,494 classifications (73.2 per cent) that come from volunteers who are logged in to the Radio Galaxy Zoo interface. For the remaining classifications by volunteers who did not establish logins, it is possible that some small fraction may have seen the subject more than once. Such duplicates are removed from the final catalogue.

For the classifications of the radio emission for each subject, determining agreement between the participants is straightforward because the sets of contours are pre-identified. The participant has the option of picking only from within this limited set, although there are additional variables depending on which counterpart galaxies they associate with the radio emission, and whether multiple radio sources are considered as belonging to the same host galaxy or from separate sources. Consensus is first measured by taking the plurality vote (over all participants) for the unique combination of radio components assigned to different sources in the image The plurality vote is the option with the highest number of total votes. For very complex subjects, it should be noted that the plurality vote may not be the option selected by the majority of the participants.

\begin{figure*}
\includegraphics[scale=0.6]{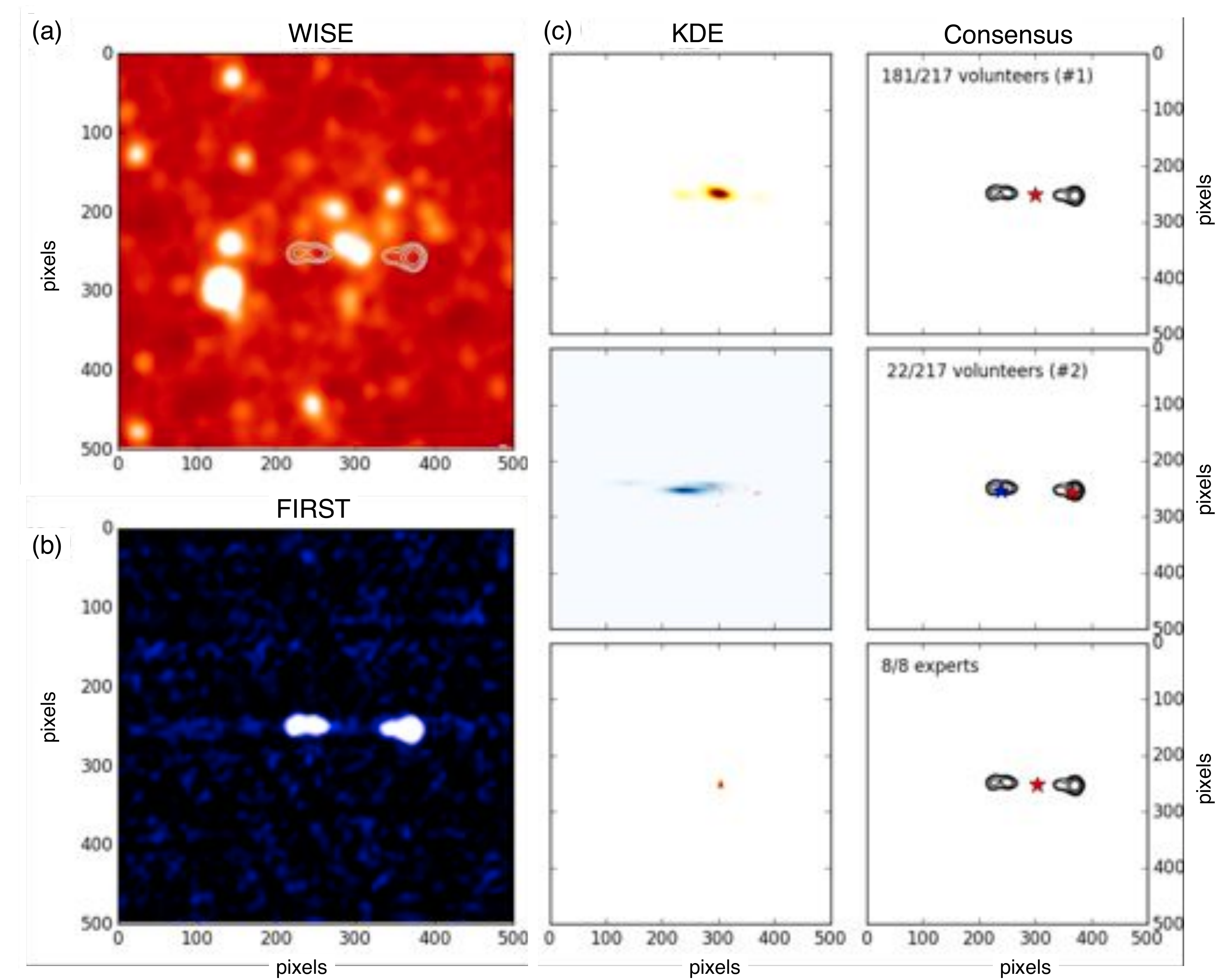}
\caption{Example of a processed RGZ subject (FIRSTJ124610.0+384838). Panel (a): $3\arcmin\times3\arcmin$ WISE 3.4~$\mu$m image. The FIRST 1.4 GHz emission is overlaid as white contours.  Panel (b): $3\arcmin\times3\arcmin$ FIRST radio continuum image. Panel (c), left column: Kernel density estimator {\bf(KDE)} used to determine the location of the IR source as pinpointed by visual identification. Panel (c), right column: Final consensus classifications, including both the FIRST radio emission components (contours) and the peak IR source or sources (stars). The top row (of panel c) shows the number one consensus classification by Radio Galaxy Zoo volunteers; the middle row (of panel c) shows the second-most common consensus among Radio Galaxy Zoo volunteers; and the bottom row (of panel c) shows the consensus of the expert Radio Galaxy Zoo science team. Both volunteers and the science team agree on the classification for this galaxy, which is of a double-lobed radio source with a single IR host at the centre. Nearest-position automated matching algorithms with a small matching radius \citep[eg, the 30\arcsec~used by][]{Kimball2008} would have split this image into two separate radio sources, corresponding to the second-most common identification by the Radio Galaxy Zoo volunteers.}
\label{fig:reduced}
\end{figure*}

The host galaxy counterpart to the radio emission is selected by the volunteer clicking on any point within the subject image (Fig.~\ref{fig:inter}b). 
Determining consensus in this case is more challenging, since the fields may be crowded. Source densities of detections in the WISE all-sky catalogue{\footnote{http://wise2.ipac.caltech.edu/docs/release/allsky/expsup/sec2\_2.html}} range from $\approx(1-2)\times10^4~\rm{deg}^{-2}$, corresponding to $25-50$ sources per $3\arcmin\times3\arcmin$ Radio Galaxy Zoo image. Using the locations of all clicks within the image, we use a kernel-density estimator (KDE) to identify the host galaxy proposed by the participants via the clustering of their click positions which may differ by a few pixels but are likely to identify the same host galaxy (see Fig.~\ref{fig:reduced}). Finally, we apply a local maximum filter to determine the number and location of detected peaks in the image, with the highest peak assigned as the location of the IR host. The only exception to this is if the plurality vote identified the radio lobes as having no visible IR counterpart; in that case, the KDE result is ignored and the consensus is assigned to ``No IR counterpart''.  In order to record the participants' clicks to a greater precision,  the pixel scale of the RGZ subjects  (as presented in Fig.~\ref{fig:reduced}) is of a higher resolution than the native pixel scales from both the FIRST and WISE images.

There is currently no weighting for individual participants in the Radio Galaxy Zoo processing.  However, we are implementing a ``gold sample'' set of 20~subjects presented to all our participants for the purpose of weighting the level of agreement between an individual participant's classification to that of a science team member. These ``gold sample'' subjects are selected to have a range of morphologies and classification difficulty, and are never removed from the broader classification pool.  The participants are unaware of the exact subjects in the ``gold sample''.  Instead, a new ``gold sample'' subject is shown to every participant at regular intervals (interspersed with the randomly-selected images) until the participant has completed the classification of all 20  ``gold sample'' subjects.  We will assemble the final Radio Galaxy Zoo catalogue using Bayesian estimators similar to those developed by \citet{Simpson2012} whereby the individual participant's classification of the ``gold sample'' will be used as seed weights for the determination of the final Bayesian classification, with the ground truth set by the science team's responses for the same subjects. 

An overview of the reduced data for an example subject is shown in Fig.~\ref{fig:reduced}.  In this particular example, we find that the cross-identifications made by the Radio Galaxy Zoo participants and the experts are consistent.  On the other hand, simple nearest-galaxy-matching algorithms \citep[e.g.,][]{Mcmahon02,Kimball2008} would classify this subject as consisting of two separate radio sources, corresponding instead to the second most-common classification made by our Radio Galaxy Zoo participants.

\section{Early results}\label{sec:science}

\subsection{WISE colours}
As an early test of the scientific returns of Radio Galaxy Zoo, we analyse the infrared colours of the host galaxies for the radio sources identified in the first twelve months of operation. It should be noted that none of the ATLAS--SWIRE subjects had been completed at this preliminary stage of the project.  From the 53,229 images with completed classifications to date, we use the raw number of votes to identify the number and association of the radio components in the image. For those radio components, we use the result from the KDE fitting to locate the position in (RA and Dec) of the infrared counterpart, if users identified one. We then match the list of positions to the WISE all-sky catalog \citep{Cutri13}. We matched 41,568 (78 per cent) of our radio sources to a WISE source within a radius of 6\arcsec. The radius is based on the size of the WISE beam at 3.4~$\mu$m; the sky density of sources out of the Galactic plane gives a mean of 0.11 random WISE sources per search cone. The majority of such spurious associations have no W2 and/or W3 emission, and are thus excluded from further analysis. The remaining IR counterparts identified by RGZ are either low $S/N$ peaks that do not pass the WISE threshold, or where the volunteers identified the radio source as having no apparent mid-IR counterpart. 

Of the Radio Galaxy Zoo sources with a WISE counterpart, we further restrict our analysis to those in which a clear identification has been made by limiting the sample to images in which at least 75 per cent of the volunteers agreed on the number and arrangement of the radio sources. This threshold is similar to the cutoffs used for the clean samples in Galaxy~Zoo \citep{Lintott2008} and Galaxy~Zoo~2 \citep{Willett2013}, but weights the sample more heavily toward single-component and/or compact sources at the expense of images with extended or multi-lobe radio morphologies. We visually inspected several hundred subjects and found reasonable agreement with this cutoff. Therefore our 75 per cent consensus sample with WISE matches consists of 33,127~sources, or 62 per cent of the classified Radio Galaxy Zoo sources to date.

Since the following analysis focuses on the infrared colour properties of galaxies, it requires a robust measurement of the infrared flux in multiple bands. We restrict the sample to those with profile $S/N \geq 5$ in $W1$, $W2$, and $W3$. It should be noted that a $S/N \geq 5$ cut translates to the WISE photometric quality class `A' and the higher $S/N$ detections of class `B' \citep[as class `B' is defined to have a $S/N \geq 3$; ][]{Cutri13}.  These comprise 100 per cent, 97 per cent, and 36 per cent, respectively, of the Radio Galaxy Zoo counterparts.  The final set of galaxies with robust RGZ identifications and clear WISE detections in three bands has a total of 4,614 galaxies.

To compare our radio-detected sample to infrared-detected sources in general, we generated a sample of $2\times10^6$ points randomly selected from sources in the WISE All-Sky Catalog located within the FIRST footprint in the northern Galactic hemisphere (RA from $10-15$~hr, dec from $0^\circ-60^\circ$). This sample is limited to the same $S/N \geq 5$ cuts as for the Radio Galaxy Zoo sources, which is roughly 5 per cent of the total WISE sample\footnote{http://wise2.ipac.caltech.edu/docs/release/prelim/expsup/sec2\_2a.html}. This sample of $\approx1\times10^5$~objects is used as a comparison control sample. 

\begin{figure*}
\includegraphics[scale=0.4]{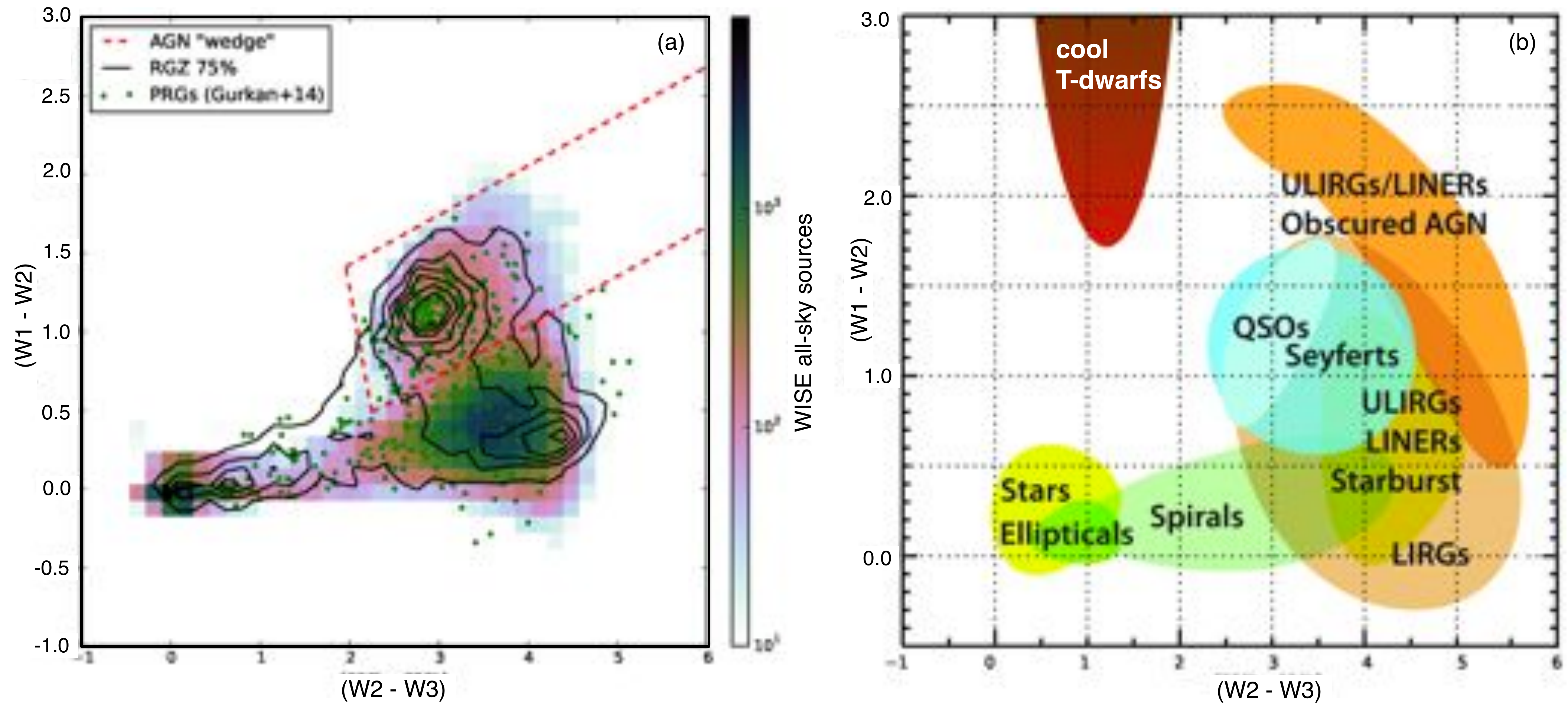}
\caption{Panel (a): WISE colour-colour diagram, showing $\sim10^5$ sources from the WISE all-sky catalog (colourmap), 4,614 sources from the 75 per cent Radio Galaxy Zoo catalogue (black contours), and powerful radio galaxies (green points) from \citet{Gurkan14}. The wedge used to identify IR colours of X-ray-bright AGN from \citet{Mateos12} is overplotted (red dashes). Only 10 per cent of the WISE all-sky sources have colours in the X-ray bright AGN wedge; this is contrasted with 40 per cent of Radio Galaxy Zoo and 49 per cent of the \citet{Gurkan14} radio galaxies. The remaining Radio Galaxy Zoo sources have WISE colours consistent with distinct populations of elliptical galaxies and LIRGs, with smaller numbers of spiral galaxies and starbursts.  Panel (b): WISE colour-colour diagram showing the locations of various classes of astrophysical objects \citep[adapted from Fig.~12 in][]{Wright2010}. }
\label{fig:wisecolours}
\end{figure*}

In Fig.~\ref{fig:wisecolours} we plot the matched WISE-RGZ sources in the infrared colour-colour space, using profile-fitted magnitudes in the $W1$, $W2$, and $W3$ bands where all WISE magnitudes are in the Vega system.  Fig.~\ref{fig:wisecolours}(a) shows the 4,614 WISE-RGZ sources from the 75 per cent consensus sample as black contours and compares our results to those from other recent studies. The underlying colourmap shows randomly selected sub-sample sources from the WISE All-Sky catalog, and the green solid points represent the 335~radio-detected galaxies sample cross-matched to WISE host galaxies by \citet{Gurkan14}. The red dashed wedge in Fig.~\ref{fig:wisecolours}(a) demarcates the infrared colour region occupied by X-ray-bright AGN \citep{Lacy04,Mateos12}.  It should be noted that the overlap between the Gurkan sample and the WISE-RGZ samples is approximately 2.3 per cent and does not significantly bias our conclusions. 

In the mid-IR bands covered by WISE, normal galaxies are expected to primarily populate a narrow mid-infrared colour band between $0.0<(W1-W2)<0.7$ and $0.5<(W2-W3)<4.0$. Since the longer $(W2-W3)$ bands are more sensitive to dust produced in star formation, spiral galaxies typically have redder colours in the mid-IR than ellipticals \citep{Wright2010}. Various classes of active galaxies (including QSOs, Seyferts, and LINERs) as well as dusty [U]LIRGs, have very red colours at longer bands $(W2-W3)>2.0$ and a broader range of colours than normal galaxies at shorter bands ($0.0<(W1-W2)<2.5$). The distribution of colours for the all-sky WISE objects spans the full range of templates for extragalactic objects shown in Fig.~\ref{fig:wisecolours}(b), but the majority of bright objects at 12~$\mu$m (W3) have colours consistent either with stars or starburst galaxies/LINERs. Consistent with recent findings \citep{Gurkan14}, the mid-infrared colour-colour plot appears to be a reasonable discriminator for many types of AGN \citep{Lacy04,Stern12,Mateos12}. It should be noted that the requirement for a detection in the $W3$-band biases our results towards low-redshift radio galaxies, as strong $W3$ emission from radio sources at high redshifts is rare.

We find that the preliminary sample of WISE-RGZ objects has a distinctly different distribution of mid-infrared colours from the randomly-selected all-sky sample. There are three primary loci. The first is at $-0.2<(W1-W2)<0.3$, $0<(W2-W3)<1$, which includes approximately 10 per cent of the Radio Galaxy Zoo sources. These colours are consistent with elliptical galaxies, which have older stellar populations and a lack of dust that results in relatively blue $(W2-W3)$ colours. The second locus of Radio Galaxy Zoo sources lies near $0.7<(W1-W2)<1.5$, $2.0<(W2-W3)<3.5$ (approximately 15 per cent of the total), corresponding to infrared colours typically associated with QSOs and Seyfert galaxies. The infrared colours are based on a strong non-thermal component from the accretion disk around the black hole. The third locus of Radio Galaxy Zoo sources lies near $0.1<(W1-W2)<0.5$, $3.5<(W2-W3)<4.8$; these are the reddest colours in $(W2-W3)$, most commonly associated with luminous infrared galaxies (LIRGs). This is the largest concentration of Radio Galaxy Zoo sources in colour-colour space, including approximately 30 per cent of Radio Galaxy Zoo sources with $W3$ measurements.

The remainder of the population of Radio Galaxy Zoo sources are distributed along the loci of both normal and active galaxies.  This is largely due to the fact that a subset of the Radio Galaxy Zoo sample consists of compact radio sources where star formation is the dominant mechanism for the observed radio emission.  The lack of objects at ($W2-W3)<0$ indicates Radio Galaxy Zoo is almost entirely free of stellar contamination.  There are also very few WISE-RGZ galaxies at the reddest $(W1-W2$) colours, indicating a lack of [U]LIRGs or very highly obscured AGN. This is consistent with results from \citet{Sajina07}, who show that ULIRGs at $z<1$ are primarily radio-quiet (although there is a larger radio-loud sample at $z\geq 2$). 

The radio-loud galaxies from the \citet{Gurkan14} sample agree with the clustering of QSO-like Radio Galaxy Zoo sources with red $(W1-W2)$ colours, although the remainder are distributed more evenly in $(W2-W3)$; their galaxies do not show the same concentration of ellipticals, and have almost no examples similar to LIRGs. Using the ``AGN wedge'' defined by \citet{Lacy04} and \citet{Mateos12}  as an AGN diagnostic, \citet{Gurkan14} find that 49 per cent of their galaxies satisfy the AGN criteria as calibrated from a bright X-ray sample (Fig.~\ref{fig:wisecolours}a). This is a powerful diagnostic for the presence of an AGN, as only 9 per cent of the WISE all-sky extragalactic sources have similar colours. However, it is clearly not a \emph{complete} sample, as more than half of their radio-loud galaxies fall outside this locus. The fraction of WISE-RGZ sources  falling within the `AGN wedge' is very similar, accounting for 40 per cent of our sample. Analysis in future papers will probe the differences between the samples, including the likely dependence on radio luminosity from brighter radio galaxies.
 
The population of Radio Galaxy Zoo host galaxies that have infrared colours consistent with massive elliptical hosts agrees with previous observations at low redshift \citep[e.g.\ ][]{auriemma77,dunlop03}. This is typically explained as the result of the accretion of smaller neighbouring galaxies, in which the resulting host is an elliptical galaxy and the radio-loud jets are launched from the recently-fueled central black hole.  To date, four examples of spiral galaxies hosting a double-lobed radio source have been discovered \citep{morganti11,hota11,bagchi14,Mao15} and Radio Galaxy Zoo has identified several such new candidates. An optical follow-up of these candidates will determine the morphology of these hosts and the relative accuracy of IR colour as a proxy. 

\begin{figure}
\includegraphics[scale=0.45]{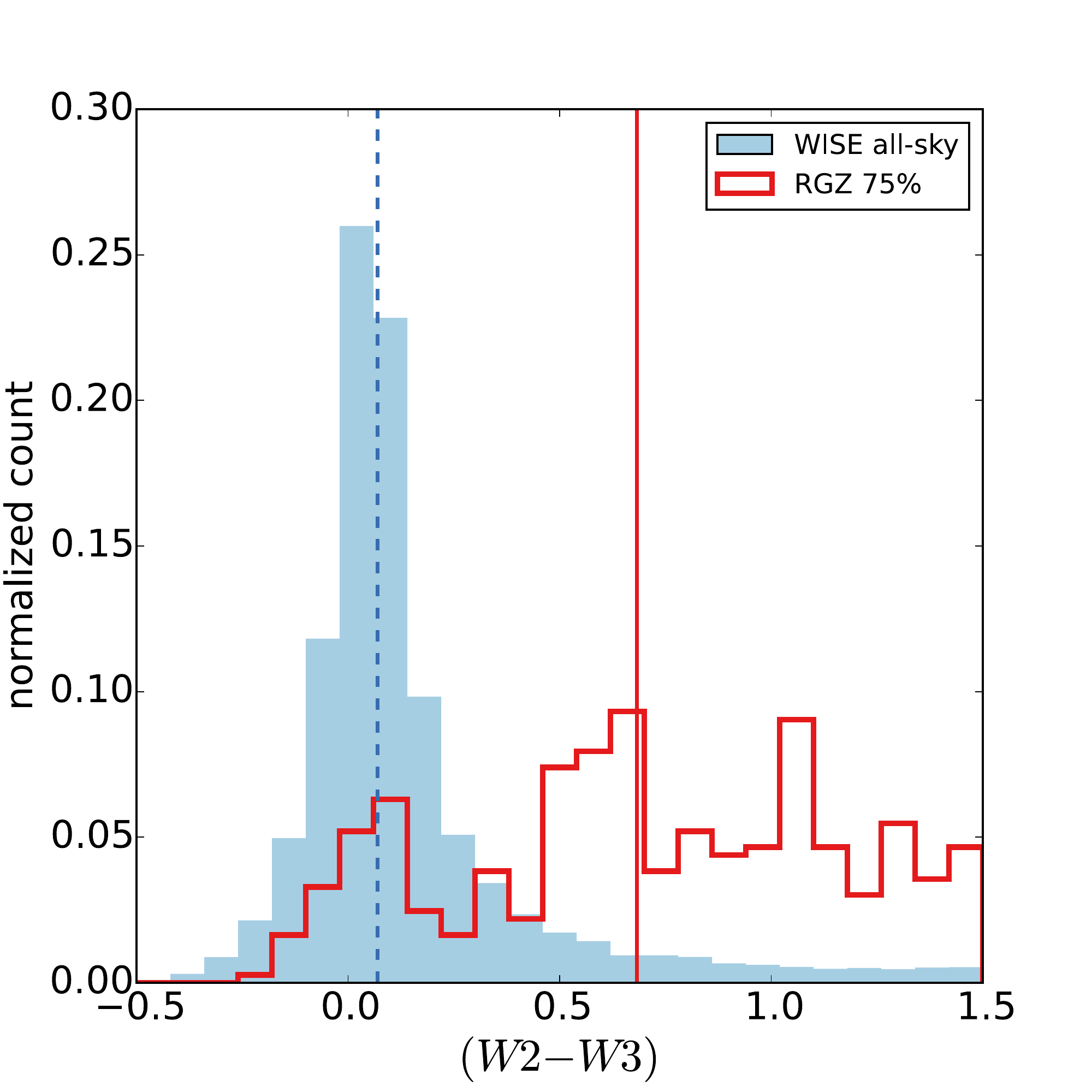}
\caption{Distribution of $(W2-W3)$ infrared colours for objects near the locus typically identified as elliptical galaxies (where $(W1-W2) < 0.5$). Solid and dashed vertical lines show the median colours of the all-sky and RGZ sources. While sources randomly selected from the WISE all-sky sample peak near $(W2-W3)=0$, our current RGZ sample shows a large population with significantly redder colours---possibly from star-forming galaxies and/or ellipticals with enhanced dust. }
\label{fig:elliptical}
\end{figure}

The distribution of sources in the elliptical region, however, is significantly different for Radio Galaxy Zoo sources vs. ``normal'' elliptical galaxies detected in the all-sky catalogue. Fig.~\ref{fig:elliptical} shows the distribution of $(W2-W3)$ for both populations. There is a clear peak for both all-sky sources and Radio Galaxy Zoo hosts around $(W2-W3)=0$.  However, the Radio Galaxy Zoo hosts have a significant population of galaxies with redder colours, out to $(W2-W3)\simeq1.5$.  Such a result suggests that the Radio Galaxy Zoo host galaxies may have enhanced dust masses over quiescent ellipticals, which would contribute to redder mid-infrared colours.  This hypothesis is consistent with previous optical studies which found that dust is prevalent in the cores of the host galaxies of 3CR radio sources \citep[e.g.\ ][]{martel99}.

On the other hand, the emission from star-forming galaxies is likely to contribute to the redder mid-infrared colours as approximately 16 per cent of the FIRST-derived Radio Galaxy Zoo sample consists of compact radio sources. However, we cannot distinguish between AGN-dominated radio emission in galaxies with on-going star formation from those galaxies where the the AGN radio emission is negligible. The peak that we find that is redder than the elliptical population may be a result of a combination of dusty ellipticals and some star-forming spirals as we have not attempted to split these.  Although \citet{Tadhunter14} find similar enhancement of dust masses for radio-loud galaxies at $0.05<z<0.7$ based on \emph{Herschel} data, a recent study by \citet{rees15} finds no difference in IR colours between radio-loud and radio-quiet elliptical host galaxies.  The properties of the Radio Galaxy Zoo elliptical population will be fully explored in a follow-up paper.

The association of radio-loud hosts with LIRGs (but not ULIRGs) is also unusual, since only a small fraction of LIRGs are associated with late-stage mergers \citep{stierwalt13}. The radio-continuum properties of 46 LIRGs from the Great Observatories All-sky LIRG Survey (GOALS) show that 45 per cent of galaxies with radio emission have radio properties resembling pure AGN, rather than starburst or starburst-AGN composites \citep{Vardoulaki15}. We note that this result is based on a sample of 46 low-redshift ($z<0.088$) LIRGs---a small fraction of the total GOALS sample of 202~galaxies. Results from Radio Galaxy Zoo, both by matching the hosts and measuring extended radio morphology vs. compact sources, can better quantify this trend as a function of redshift. 

The clustering of radio-detected WISE counterparts in all three loci (ellipticals, QSOs, and LIRGs) and their difference from random all-sky WISE sources strongly implies that Radio Galaxy Zoo classifiers are accurately matching the radio lobes to their host galaxies. Spurious associations would result in infrared colours which are more consistent with stars or starburst galaxies. These early results (which have not been subject to explicit user weighting or outlier rejection) reinforce the ability of crowdsourced volunteers to carry out tasks useful for astronomical research in a reliable manner.

\subsection{New discoveries through RadioTalk}

\begin{figure*}
\includegraphics[scale=.85]{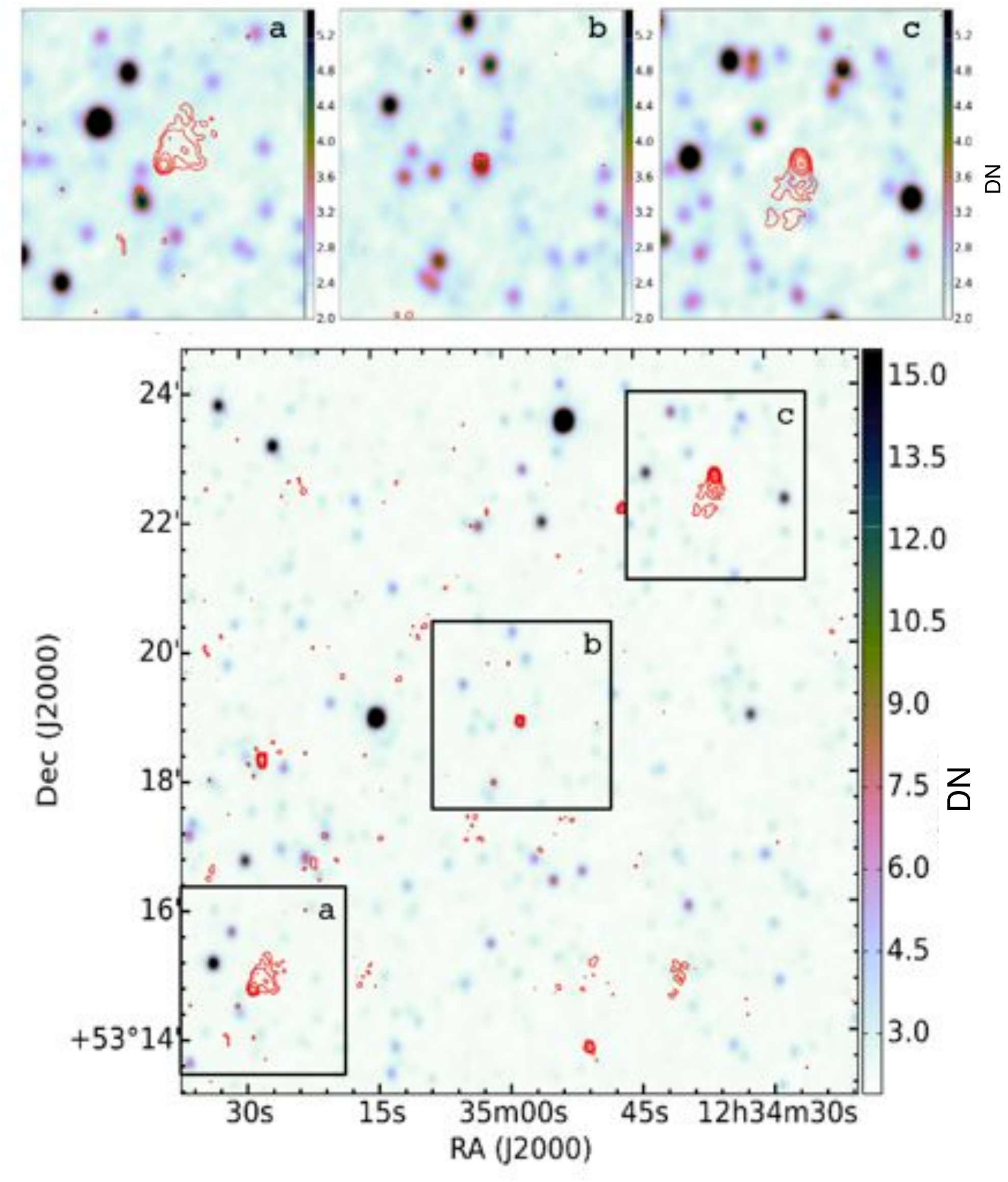}
\caption{An example of how some of the volunteers recognise that they might be looking at only a piece of a radio source, and then use the provided links in RadioTalk to examine larger fields and other surveys.  The three small insets labelled A, B, and C are $3\arcmin \times 3\arcmin$ in size representing the Radio Galaxy Zoo images presented to the participants.  The much larger field ($11.5\arcmin \times 11.5\arcmin$) shows that this is part of a very large radio triple, with a $670\arcsec$ angular size from hot spot A to hot spot C.   The background image is the WISE mid-infrared image and the contours show the FIRST radio data with the contours starting at 3 times the local rms (0.14~mJy~beam$^{-1}$) and increasing by a factor of 2.  The background colour scheme comes from {\tt CUBEHELIX} \citep{Green2011}.}
\label{fig:unexpected}
\end{figure*}

The most beneficial features of the RadioTalk online forum are: (1) the links to larger fields and other complementary surveys; and (2) the discussion board.  Fig.~\ref{fig:unexpected} illustrates one example of the power of RadioTalk.  The Radio Galaxy Zoo image size presented to the participants is $3\arcmin \times 3\arcmin$ and three such squares are shown in Fig.~\ref{fig:unexpected}a -- c.  Using the tools provided in RadioTalk, it became apparent to several Radio Galaxy Zoo participants that the radio components observed in these three subjects are part of the same radio source extending $11.1\arcmin$ in angular size (large panel in Fig.~\ref{fig:unexpected}).  The optical host galaxy is SDSS~J123458.46+531851.3 and has a photometric redshift of $z=0.62 \pm 0.1$. This source had been found in an independent visual search for giant radio sources in the NVSS (Andernach et al. 2012). The overall radio size of 4.6 Mpc makes it the third-largest radio galaxy
known (Andernach, priv. comm. 2014). Given the presence of the unrelated radio and infrared sources in this field, only a visual inspection would allow the identification of this triple radio source. 

\begin{figure}
\includegraphics[scale=.35]{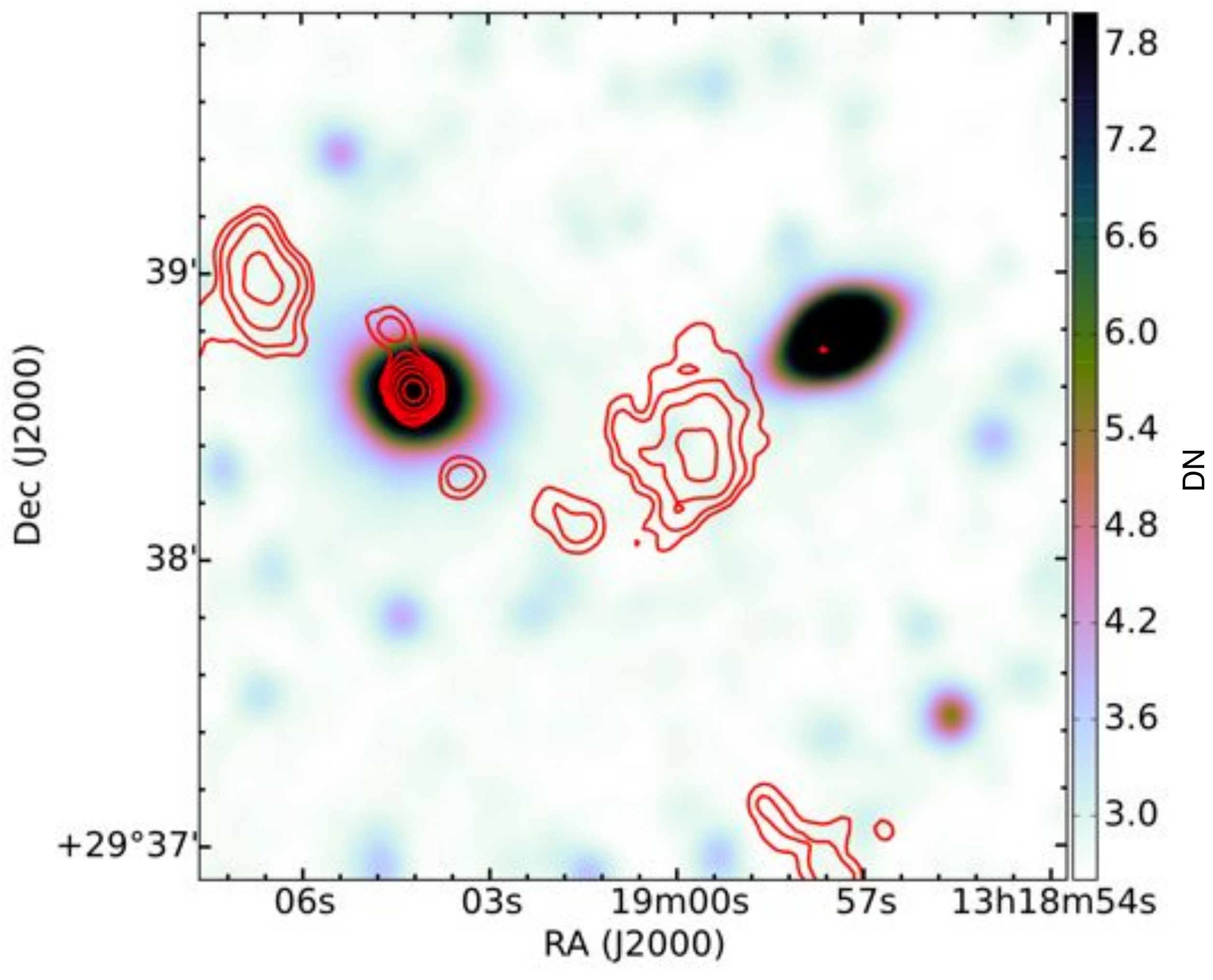}
\caption{An example of a galaxy where  visual identification of the radio components is necessary. The automated algorithms would have classified the non-core emission as independent sources, whereas the Radio Galaxy Zoo volunteers (in agreement with the experts) find all five radio emission components in the upper half of the image to be related to the same source. }
\label{fig:unexpected2}
\end{figure}

Even for radio sources much less extended than the one presented in Fig.~\ref{fig:unexpected},  automated algorithms based on: (1) nearest position matching \citep[e.g.\ ][]{Mcmahon02,Kimball2008}; or (2) a combination of position matching with a specific search for double-lobes \citep[e.g.\ ][]{vanvelzen15} can be confused by the presence of multiple discrete components typical of non-compact radio sources.  Fig.~\ref{fig:unexpected2} shows an example of a radio source with a bent, double-lobed morphology in a galaxy group at $z=0.073$. Apart from the radio emission from the core, an automated algorithm will have difficulty in determining whether the discrete components are lobes belonging to the core or if they are independent sources.  On the other hand, there is strong agreement between the visual classifications by the RGZ volunteers and the experts that all the visible radio components are part of the same bent radio source structure hosted by the galaxy, SDSS~J131904.16+293834.8.

The discovery from RadioTalk of a re-started jet in a WAT found within a few days of the Radio Galaxy Zoo launch was unexpected. We have since conducted follow-up spectroscopic observations to determine the redshift of the object, as well as deeper radio continuum observations with the VLA.  

In addition to unexpected discoveries, we also have ongoing collaborations between the scientists and the Radio Galaxy Zoo volunteers on various research topics. Typically, the scientists will communicate directly with the Radio Galaxy Zoo volunteers by explaining their interests in a particular object or phenomenon and then request help in collating lists of possible candidates from the objects that have been inspected.  Currently, the projects being facilitated by RadioTalk include: (1) the search for hybrid radio sources where one radio source appears to have both FRI and FRII characteristics (known as HyMoRS; Kapi\'{n}ska et al. (2015) submitted to MNRAS); (2) the search for double-lobed radio sources associated with spiral host galaxies (led by Mao); and (3) the identification of giant radio galaxies (led by Andernach). 

\section{Summary}\label{sec:summary}
Radio Galaxy Zoo is an online citizen science project operating within the Zooniverse initiative where volunteers can contribute towards current research projects.  The primary purpose of Radio Galaxy Zoo is to obtain host identifications for radio sources from wide-field and eventually all-sky radio surveys.  In preparation for the next generation of all-sky radio surveys, such as EMU which will yield 70 million sources, we are also testing the viability of citizen science as an alternative technique for inspecting such large datasets.  In its first and current incarnation launched publicly in December 2013, we are cross-matching the FIRST and ATLAS radio surveys to mid-infrared images from the WISE and SWIRE surveys.

By combining the work of more than 4,000 participants in the first 12 months of operation, we have obtained more than 30,000 host identifications from Radio Galaxy Zoo with greater than 75 per cent consensus. By matching these to nearby WISE detections, we find that the majority of our current sample of radio sources reside in mid-infrared colour-colour regions that are known to be occupied by elliptical galaxies, QSOs, and LIRGs. This result is consistent with canonical understanding whereby radio-loud sources are primarily affiliated with elliptical galaxies and late-stage mergers. We also find a significant population of Radio Galaxy Zoo sources with redder mid-infrared colours than normal elliptical galaxies.  This is either IR emission from star-forming galaxies or evidence of enhanced dust content. Further analysis will examine how the association with the host depends on radio morphology and power. 

While we still have a significant population of sources yet to be quantified ($>80$ per cent), we do find that the project participants are as effective as the science team at identifying host galaxies for sources which are currently too complex (due to a combination of structures and/or number of source components) for simple position-matching automated algorithms.  In addition, the experienced participants are also very successful at the identification of radio source candidates which extend beyond the given $3\arcmin \times 3\arcmin$ field.  However, it should be noted there remains a significant number of radio sources at the 10 -- 20 per cent level which are too complex to allow an unambiguous identification of the host without further follow-up observations.

Additionally, through the collaborative efforts between participants and the science team, we have discovered multiple examples of unusual radio galaxies, including spiral galaxies with extended double-lobed radio emission and new HyMoRS.  

\section*{Acknowledgments}
This publication has been made possible by the participation of more than 6900 volunteers in the Radio Galaxy Zoo project.  Their contributions are individually acknowledged at http://rgzauthors.galaxyzoo.org. We thank our referee, P. Leahy for his constructive comments which helped improve this paper.  We also acknowledge A.~Kapadia, A.~Smith, M.~Gendre, and S.~George who have made contributions to the project.  OIW acknowledges a Super Science Fellowship from the Australian Research Council.  Partial support for this work for KW, LR and AG is provided by the U.S.\ National Science Foundation grant AST-112595 to the University of Minnesota.   SS thanks the Australian Research Council for an Early Career Fellowship DE130101399.  NS is the recipient of an Australian Research Council Future Fellowship.  JKB and ADK acknowledges funding from the Australian Research Council Centre of Excellence for All-sky Astrophysics (CAASTRO), through project number CE110001020.  KS gratefully acknowledges support from Swiss National Science Foundation Professorship grant PP00P2\_138979/1.

This publication makes use of data products from the {\it Wide-field Infrared Survey Explorer} and the {\it Spitzer Space Telescope}.  The {\it Wide-field Infrared Survey Explorer} is a joint project of the University of California, Los Angeles, and the Jet Propulsion Laboratory/California Institute of Technology, funded by the National Aeronautics and Space Administration.  SWIRE is supported by NASA through the SIRTF Legacy Program under contract 1407 with the Jet Propulsion Laboratory.  This publication makes use of radio data from the Australia Telescope Compact Array and the Karl G. Jansky Very Large Array (operated by NRAO).   The Australia Telescope Compact Array is part of the Australia Telescope, which is funded by the Commonwealth of Australia for operation as a National Facility managed by CSIRO.   The National Radio Astronomy Observatory is a facility of the National Science Foundation operated under cooperative agreement by Associated Universities, Inc.


\bibliographystyle{mn2e}

\begin{thebibliography}{}
\bibitem[Ahn et al. (2014)]{SDSS10} Ahn, C.~P., Alexandroff, R., Allende Prieto, C., et al.\ 2014, \apjs, 211, 17 
\bibitem[Alam et al.(2015)]{Alam15} Alam, S., Albareti, F.~D., Allende Prieto, C., et al.\ 2015, arXiv:1501.00963 
\bibitem[\protect\citeauthoryear{Andernach \& al.}{Andernach et~al.}{2012}]{Andernach2012} Andernach H., Jim\'enez Andrade E.~F., Maldonado S\'anchez R.~F., V\'asquez Baez I.~R.,  2012, in ``Science from the Next Generation Imaging and Spectroscopic Surveys'', see http://adsabs.harvard.edu/abs/2012sngi.confP...1A
\bibitem[\protect\citeauthoryear{Auriemma, Perola, Ekers \& al}{Auriemma et~al.}{1977}]{auriemma77} Auriemma, C., Perola, G.~C., Ekers, R.~D., et al.\ 1977, A\&A, 57, 41 
\bibitem[Bagchi et al.(2014)]{bagchi14} Bagchi, J., Vivek, M., Vikram, V., et al.\ 2014, \apj, 788, 174 
\bibitem[Becker et al. (1995)]{Becker95}Becker, R. H., White, R. L., \& Helfand, D. J. 1995, \apj, 450, 559
\bibitem[Best(2009)]{best09} Best, P.~N.\ 2009, Astronomische Nachrichten, 330, 184 
\bibitem[Bridle et al.(1989)]{Bridle89} Bridle, A.~H., Fomalont, E.~B., Byrd, G.~G., \& Valtonen, M.~J.\ 1989, \aj, 97, 674 
\bibitem[Burns(1998)]{Burns98} Burns, J.~O.\ 1998, Science, 280, 400 
\bibitem[Cardamone et al. (2009)]{Cardamone09} Cardamone, C., Schawinski, K., Sarzi, M., et al.\ 2009, \mnras, 399, 1191 
\bibitem[Condon et al. (1990)]{condon90} Condon, J.~J., Helou, G., Sanders, D.~B., \& Soifer, B.~T. 1990, \apjs, 73, 359
\bibitem[Condon et al. (1991)]{condon91} Condon, J.~J., Huang, Z.-P., Yin, Q.~F., \& Thuan, T.~X. 1991, \apj, 378, 65
\bibitem[\protect\citeauthoryear{Condon, Cotton, Greisen, Yin, Perley, Taylor \& Broderick}{Condon et~al.}{1998}]{Condon1998} Condon J.~J.,  Cotton W.~D.,  Greisen E.~W.,  Yin Q.~F.,  Perley R.~A.,  Taylor G.~B.,    Broderick J.~J.,  1998, AJ, 115, 1693
\bibitem[\protect\citeauthoryear{Condon, Cotton, Fomalont, Kellermann, Miller, Perley, Scott, Vernstrom \& Wall}{Condon et~al.}{2012}]{condon12}Condon, J.~J. et al., 2012, \apj, 758, 23
\bibitem[Cutri \& et al.(2013)]{Cutri13} Cutri, R.~M., Wright, E.~L., Conrow, T. \& al.\ 2013, VizieR Online Data Catalog, 2328, 0 
\bibitem[\protect\citeauthoryear{Dunlop, McLure, Kukula \& al.}{Dunlop et~al.}{2003}]{dunlop03} Dunlop, J.~S., McLure, R.~J., Kukula, M.~J., et al.\ 2003, \mnras, 340, 1095 
\bibitem[Eilek et al.(1984)]{Eilek84} Eilek, J.~A., Burns, J.~O., O'Dea, C.~P., \& Owen, F.~N.\ 1984, \apj, 278, 37 
\bibitem[Fan et al.(2015)]{Fan15}Fan, D., Budav{\'a}ri, T., Norris, R.~P., \&  Hopkins, A.~M.\ 2015, \mnras, in press. (arXiv:1505.00621)
\bibitem[\protect\citeauthoryear{Fanaroff \& Riley}{Fanaroff \&  Riley}{1974}]{Fanaroff1974} Fanaroff B.~L.,  Riley J.~M.,  1974, \mnras, 167, 31P
\bibitem[Fortson et al.(2012)]{Fortson12} Fortson, L., Masters, K., Nichol, R., et al.\ 2012, Advances in Machine Learning and Data Mining for Astronomy, CRC Press, Taylor \& Francis Group, eds.: Michael J. Way, Jeffrey D. Scargle, Kamal M. Ali, Ashok N. Srivastava, 213 
\bibitem[\protect\citeauthoryear{Franzen \& al}{Franzen et~al.}{2015}]{Franzen15}Franzen, T.~M.~O., Banfield, J.~K. et al., 2015, \mnras, submitted
\bibitem[\protect\citeauthoryear{Gendre, Best \& Wall}{Gendre et~al.}{2010}]{Gendre2010} Gendre M.~A.,  Best P.~N.,    Wall J.~V.,  2010, \mnras, 404, 1719
\bibitem[\protect\citeauthoryear{Gendre, Best, Wall \& Ker}{Gendre et~al.}{2013}]{Gendre2013} Gendre M.~A.,  Best P.~N.,  Wall J.~V.,    Ker L.~M.,  2013, \mnras, 430, 3086
\bibitem[\protect\citeauthoryear{{Gopal-Krishna} \& Wiita}{{Gopal-Krishna} \&  Wiita}{2000}]{Gopal2000} {Gopal-Krishna} \& Wiita P.~J.,  2000, A\&A, 363, 507
\bibitem[\protect\citeauthoryear{Green}{Green}{2011}]{Green2011}Green D.~A.,  2011, BASI, 39, 289
\bibitem[\protect\citeauthoryear{G{\"u}rkan, Hardcastle \& Jarvis}{G{\"u}rkan et~al.}{2014}]{Gurkan14} G{\"u}rkan, G., Hardcastle, M.~J., \& Jarvis, M.~J.\ 2014, \mnras, 438, 1149 
\bibitem[Hota et al. (2011)]{hota11}Hota, A., Sirothia, S.~K., Ohyama, Y. et al. 2011, \mnras, 417, L36
\bibitem[Jarvis(2012)]{Jarvis12} Jarvis, M.~J.\ 2012, African Skies, 16, 44 
\bibitem[\protect\citeauthoryear{Johnston, Bailes, Bartel, Baugh, Bietenholz, Blake, Braun \& al}{Johnston et~al.}{2007}]{Johnston2007}Johnston S., Bailes, M., Bartel, N. et al.,  2007, PASA, 24, 174
\bibitem[\protect\citeauthoryear{Jonas}{Jonas}{2009}]{Jonas2009} Jonas J.~L.,  2009, in Proceedings of the IEEE {MeerKAT - The South African Array With Composite Dishes and Wide-Band Single Pixel Feeds}. pp 1522--1530
\bibitem[\protect\citeauthoryear{Keel, Lintott, Schawinski, Bennert, Thomas, Manning, Chojnowski, van Arkel \& Lynn}{Keel et~al.}{2012}]{Keel2012} Keel W.~C., Lintott, C.~J., Schawinski, K. et al.,  2012, AJ, 144, 66
\bibitem[Kimball \& Ivezi{\'c}(2008)]{Kimball2008} Kimball, A.~E., \& Ivezi{\'c}, {\v Z}.\ 2008, \aj, 136, 684 
\bibitem[Lacy et al.(2004)]{Lacy04} Lacy, M., Storrie-Lombardi, L.~J., Sajina, A., et al.\ 2004, \apjs, 154, 166 
\bibitem[Laing(1996)]{Laing96} Laing, R.~A.\ 1996, Energy Transport in Radio Galaxies and Quasars, eds. P. E. Hardee, A. H. Bridle, and J. A. Zensus, ASP Conf. Ser. 100, San Francisco: Astronomical Society of the Pacific, 241 
\bibitem[Leahy et al.(1996)]{Leahy1996} Leahy, J.~P., Bridle, A.~H., \& Strom, R.~G.\ 1996, Extragalactic radio sources, IAU Symp. 175, 157, eds.  R.D. Ekers, C. Fanti, and L. Padrielli, Kluwer Academic Publishers
\bibitem[Lin et al. (2010)]{Lin10} Lin, Y.-T., Shen, Y., Strauss, M.~A., Richards, G.~T., \& Lunnan, R.\ 2010, \apj, 723, 1119 
\bibitem[\protect\citeauthoryear{Lintott, Schawinski, Slosar, L and, Bamford, Thomas, Raddick, Nichol, Szalay, Andreescu, Murray \& Vandenberg}{Lintott et~al.}{2008}]{Lintott2008} Lintott C.~J., Schawinski, K., Slosar, A. et al.,  2008, \mnras, 389, 1179
\bibitem[\protect\citeauthoryear{Lintott, Schawinski, Keel, van Arkel, Bennert,  Edmondson, Thomas, Smith, Herbert, Jarvis, Virani, Andreescu, Bamford, Land,  Murray, Nichol, Raddick, Slosar, Szalay \& Vandenberg}{Lintott et~al.}{2009}]{Lintott2009} Lintott C.~J., Schawinski, K., Keel, W.~C. et al.,  2009, \mnras, 399, 129
\bibitem[\protect\citeauthoryear{Lonsdale, Smith, Rowan-Robinson, Surace,  Shupe, Xu \& Oliver}{Lonsdale et~al.}{2003}]{Lonsdale2003}Lonsdale C.~J., Smith, H.~E., Rowan-Robinson, M. et al.,  2003, PASP, 115, 897
\bibitem[\protect\citeauthoryear{Mackay}{Mackay}{1971}]{Mackay1971} Mackay C.~D.,  1971, \mnras, 154,  209
\bibitem[Mao et al.(2012)]{Mao12} Mao, M.~Y., Sharp, R., Norris, R.~P., et al.\ 2012, \mnras, 426, 3334 
\bibitem[Mao et al.(2015)]{Mao15} Mao, M.~Y., Owen, F., Duffin, R., et al.\ 2015, \mnras, 446, 4176 
\bibitem[Martel et al.(1999)]{martel99} Martel, A.~R., Baum, S.~A., Sparks, W.~B., et al.\ 1999, \apjs, 122, 81 
\bibitem[\protect\citeauthoryear{Mauch \& Sadler}{Mauch \&  Sadler}{2007}]{Mauch2007}Mauch T.,  Sadler E.~M.,  2007, \mnras, 375, 931
\bibitem[\protect\citeauthoryear{Mateos, Alonso-Herrero, Carrera \& al}{Mateos et~al.}{2012}]{Mateos12}Mateos, S., Alonso-Herrero, A., Carrera, F.~J., et al.\ 2012, \mnras, 426, 3271 
\bibitem[McMahon et al.(2002)]{Mcmahon02} McMahon, R.~G., White, R.~L., Helfand, D.~J., \& Becker, R.~H.\ 2002, \apjs, 143, 1 
\bibitem[Middelberg et al. (2008)]{middelberg08}Middelberg, E., Norris, R.~P., Cornwell, T.~J. et al., 2008, \aj, 135, 1276 
\bibitem[Morganti et al. (2011)]{morganti11} Morganti, R., Holt, J., Tadhunter, C. et al.\ 2011, \aap, 535, A97
\bibitem[Norris et al. (2006)]{Norris06}Norris, R.P., et al. 2006, \aj, 132, 2409 
\bibitem[\protect\citeauthoryear{Norris, Hopkins, Afonso, Brown, Condon, Dunne,  Feain \& al}{Norris et~al.}{2011}]{Norris2011}Norris R.~P., Hopkins, A.~M., Afonso, J. et al.,  2011, \pasa, 28, 215
\bibitem[\protect\citeauthoryear{Owen \& Ledlow}{Owen \&  Ledlow}{1994}]{Owen1994} Owen F.~N.,  Ledlow M.~J.,  1994, The First Stromlo Symposium: The Physics of Active Galaxies. ASP Conference Series, 54, 319
\bibitem[\protect\citeauthoryear{Owen \& Morrison}{Owen \&  Morrison}{2008}]{Owen2008} Owen F.~N.,  Morrison G.~E.,  2008, AJ, 136, 1889
\bibitem[\protect\citeauthoryear{Owen \& Rudnick}{Owen \&  Rudnick}{1976}]{Owen1976} Owen F.~N.,  Rudnick L.,  1976, \apj, 205, L1
\bibitem[Proctor(2006)]{Proctor06} Proctor, D.~D.\ 2006, \apjs, 165, 95 
\bibitem[Proctor(2011)]{Proctor11} Proctor, D.~D.\ 2011, \apjs, 194, 31 
\bibitem[Rees et al.\ (2015)]{rees15} Rees, G.~A., Spitler, L.~R., Norris, R.~P. et al., 2015, \mnras, submitted
\bibitem[R\"ottgering et al. (2011)]{Rottgering11}R\"ottgering, H., Afonso, J., Barthel, P. et al., 2011, \japa, 32, 557
\bibitem[\protect\citeauthoryear{Rudnick \& Owen}{Rudnick \& Owen}{1976}]{Rudnick1976} Rudnick L.,  Owen F.~N.,  1976, \apj, 203, L107
\bibitem[Sadler et al.(2014)]{Sadler14} Sadler, E.~M., Ekers, R.~D., Mahony, E.~K., Mauch, T., \& Murphy, T.\ 2014, \mnras, 438, 796 
\bibitem[\protect\citeauthoryear{Sajina, Yan, Armus \& al}{Sajina et~al.}{2007}]{Sajina07} Sajina, A., Yan, L., Armus, L., et al.\ 2007, \apj, 664, 713 
\bibitem[Seymour et al. (2008)]{Seymour2008}Seymour, N., Dwelly, T., Moss, D. et al. 2008, \mnras, 386, 1695
\bibitem[Seymour(2009)]{Seymour2009} Seymour, N.\ 2009, Panoramic Radio Astronomy: Wide-field 1-2 GHz Research on Galaxy Evolution, Proceedings of Science, PoS(PRA2009)035, eds. G.Heald \& P.Serra 
\bibitem[Shabala et al. (2008)]{Shabala08} Shabala, S.~S., Ash, S., Alexander, P., \& Riley, J.~M.\ 2008, \mnras, 388, 625 
\bibitem[\protect\citeauthoryear{Shabala \& Alexander}{Shabala \& Alexander}{2009}]{Shabala2009} Shabala S. \&  Alexander P.,  2009, \apj, 699, 525
\bibitem[Simpson et al. (2012)]{Simpson2012} Simpson, E., Roberts, S., Psorakis, I, \& Smith, A.\ 2012, arXiv:1206.1831
\bibitem[Slee et al.(1994)]{Slee94} Slee, O.~B., Sadler, E.~M., Reynolds, J.~E., \& Ekers, R.~D.\ 1994, \mnras, 269, 928 
\bibitem[\protect\citeauthoryear{Smol{\v c}i{\'c}, Schinnerer, Zamorani, Bell, Bondi, Carilli, Ciliegi, Mobasher, Paglione, Scodeggio \& Scoville}{Smol{\v c}i{\'c} et~al.}{2009}]{Smolcic2009} Smol{\v c}i{\'c}, V.,  Schinnerer, E., Zamorani, G. et al., 2009, \apj, 690, 610S
\bibitem[\protect\citeauthoryear{Stern, Assef, Benford \& al}{Stern et~al.}{2012}]{Stern12} Stern, D., Assef, R.~J., Benford, D.~J., et al.\ 2012, \apj, 753, 30 
\bibitem[\protect\citeauthoryear{Stierwalt, Armus, Surace \& al}{Stierwalt et~al.}{2013}]{stierwalt13} Stierwalt, S., Armus, L., Surace, J.~A., et al.\ 2013, \apjs, 206, 1 
\bibitem[\protect\citeauthoryear{Swain, Bridle \& Baum}{Swain et~al.}{1998}]{Swain1998}Swain M.~R.,  Bridle A.~H.,    Baum S.~A.,  1998, \apj,  507, L29
\bibitem[Tadhunter et al.(2014)]{Tadhunter14} Tadhunter, C., Dicken, D., Morganti, R., et al.\ 2014, \mnras, 445, L51 
\bibitem[van Velzen et al.(2015)]{vanvelzen15} van Velzen, S., Falcke, H., K\"ording, E.\ 2015, \mnras, 446, 2985 
\bibitem[Vardoulaki et al.(2015)]{Vardoulaki15} Vardoulaki, E., Charmandaris, V., Murphy, E.~J., et al.\ 2015, \aap, 574, A4 
\bibitem[Venkatesan et al.(1994)]{Venkatesan94} Venkatesan, T.~C.~A., Batuski, D.~J., Hanisch, R.~J., \& Burns, J.~O.\ 1994, \apj, 436, 67 
\bibitem[Verheijen et al.(2008)]{Verheijen08} Verheijen, M.~A.~W., Oosterloo, T.~A., van Cappellen, W.~A., et al.\ 2008, The Evolution of Galaxies Through the Neutral Hydrogen Window, AIP Conference Proceedings, Volume 1035, p. 265 
\bibitem[\protect\citeauthoryear{White, Becker, Helfand \& Gregg}{White et~al.}{1997}]{White1997} White R.~L.,  Becker R.~H.,  Helfand D.~J.,    Gregg M.~D.,  1997, \apj, 475, 479
\bibitem[\protect\citeauthoryear{Willett, Lintott, Bamford \& al}{Willett et~al.}{2013}]{Willett2013} Willett, K.~W., Lintott, C.~J., Bamford, S.~P., et al.\ 2013, \mnras, 435, 2835 
\bibitem[\protect\citeauthoryear{Wright, Eisenhardt, Mainzer, Ressler, Cutri \& Jarrett}{Wright et~al.}{2010}]{Wright2010} Wright E.~L., Eisenhardt, P.~R.~M., Mainzer, A.~K. et al.,  2010, AJ, 140, 1868

\end{thebibliography}

\label{lastpage}

\end{document}